\documentclass[a4paper, onecolumn, 10pt]{iopart}

\usepackage{epsfig,subfigure,graphicx,amssymb,amscd,amsfonts}

\usepackage{cite}

\usepackage[table]{xcolor}
\usepackage{hyperref}

\usepackage{marginnote,ifthen}
\newcommand\havecomments{1} 

\newcommand\rightnote[1]{\ifthenelse{\isodd{\havecomments}}{\normalmarginpar\marginnote{\footnotesize \textcolor{violet}{\textit{[#1]}}}}{}}
\newcommand\leftnote[1]{\ifthenelse{\isodd{\havecomments}}{\reversemarginpar\marginnote{\footnotesize \textcolor{violet}{\textit{[#1]}}}}{}}

\newcommand\eg{\emph{e.g.}~}
\newcommand\ie{\emph{i.e.}~}

\newcommand\figref[1]{Fig.~\ref{#1}}
\newcommand\eqref[1]{(\ref{#1})}

\renewcommand{\i}[0]{\imath}

\newcommand{\ket}[1]{\left|#1\right\rangle}

\begin{document}

\title{Dynamics and level statistics of interacting fermions in the Lowest Landau Level}

\author{
M.~Fremling,$^1$
C\'ecile Repellin,$^{2,3}$
Jean-Marie St\'ephan,$^{2,4}$
N.~Moran,$^1$
J.~K.~Slingerland,$^{1,5}$
and Masudul Haque$^{1,2}$}

\address{$^1$ Department of Theoretical Physics, Maynooth University, Co. Kildare, Ireland}
\address{$^2$ Max Planck Institute for the Physics of Complex Systems, N{\"o}thnitzer Str.~38, 01187 Dresden, Germany}
\address{$^3$ Department of Physics, Massachusetts Institute of Technology, Cambridge, MA 02139, USA}
\address{$^4$ Univ Lyon, CNRS, Universit\'e Claude Bernard Lyon 1, UMR5208, Institut Camille Jordan, F-69622 Villeurbanne, France}
\address{$^5$ Dublin Institute for Advanced Studies, School of Theoretical Physics, 10 Burlington Rd, Dublin, Ireland}

\date{\today}

\begin{abstract}

  We consider the unitary dynamics of interacting fermions in the lowest Landau level, on spherical
  and toroidal geometries.  The dynamics are driven by the interaction Hamiltonian which, viewed in
  the basis of single-particle Landau orbitals, contains correlated pair hopping terms in addition
  to static repulsion.  This setting and this type of Hamiltonian has a significant history in
  numerical studies of fractional quantum Hall (FQH) physics, but the many-body quantum dynamics
  generated by such correlated hopping has not been explored in detail.  We focus on initial states
  containing all the fermions in one block of orbitals.  We characterize in detail how the fermionic
  liquid spreads out starting from such a state.  We identify and explain differences with regular
  (single-particle) hopping Hamiltonians.  Such differences are seen, \eg in the entanglement
  dynamics, in that some initial block states are frozen or near-frozen, and in density gradients
  persisting in long-time equilibrated states.  Examining the level spacing statistics, we show that
  the most common Hamiltonians used in FQH physics are not integrable, and explain that GOE
  statistics (level statistics corresponding to the Gaussian Orthogonal Ensemble) can appear in many
  cases despite the lack of time-reversal symmetry.

\end{abstract}

\maketitle

\section{Introduction}

In recent years, intense research effort has been directed at the real-time dynamics of many-body
quantum systems in the absence of external baths or dissipation mechanisms
\cite{PolkovnikovSilva_RevModPhys2011, Eisert_NatPhys2015, PolkovnikovRigol_AdvPhys2016, Mitra_review2018}.
Motivated by experiments --- most prominently with laser-cooled atoms,
but increasingly also in other settings ---
where experimental measurements are performed at timescales shorter than dissipation timescales,
a large and growing body of theoretical work has addressed issues such as relaxation,
thermalization, and dissipationless transport in isolated quantum many-body systems.

The bulk of the literature on non-dissipative many-body quantum dynamics focuses on short-range
Hamiltonians with density-density interactions and single-particle hopping.  In particular, the
dynamics of one-dimensional (and to a lesser extent two-dimensional) lattice Hamiltonians have
enjoyed much attention.  Models that can be mapped to free-fermion systems, such as the
transverse-field Ising model and hard-core bosons on 1D chains, as well as models that are
fundamentally interacting, such as the Heisenberg spin-$\frac{1}{2}$ chain and the bosonic and
fermionic Hubbard models, have served as widely used test beds for understanding non-equilibrium
quantum physics in the dissipationless regime \cite{PolkovnikovSilva_RevModPhys2011,
  Eisert_NatPhys2015, PolkovnikovRigol_AdvPhys2016, Mitra_review2018}.

In the present work, we consider the dynamics of a Hamiltonian which also has a venerable history in
the condensed matter physics literature, but has some remarkable properties which stand out from the
Hamiltonians more usually considered in the non-equilibrium literature.  This Hamiltonian stems from
the community studying fractional quantum Hall (FQH) physics.  We consider interacting fermions on a
two dimensional surface subject to a strong magnetic field, such that the dynamics can be considered
to be confined to the lowest Landau level (LLL) at finite filling.  Following a strong tradition in
numerical studies of FQH physics, the fermions are placed on closed geometries, either a sphere
\cite{Haldane_PRL1983, HaldaneRezayi_PRL1985, Haldane_inPrangeGirvinbook_1987,
  FanoOrtolaniColombo_PRB1986, HaldaneRezayi_PRL1988, Jain_2007_Book} or a torus
\cite{Yoshioka_1983, Haldane_PRL1985, HaldaneRezayi_PRB1985, Jain_2007_Book}.  The single-particle
basis then consists of a finite number of orbitals circling the sphere (along lines of latitude) or
circling one direction of the torus.  When the interaction potential is projected onto the LLL, we
obtain a quartic term of type $a_{i+k+m}^\dagger a_{i-m}^\dagger a_{i+k} a_{i}$, where the
subscripts are orbital indices.  This contains both static density-density repulsion ($m=0$ or
$m=-k$) and, more interestingly, correlated hoppings of particle pairs.  When focusing on the LLL,
the kinetic energy (quadratic) part of the Hamiltonian is quenched.  Thus the dynamics are driven by
the correlated pair hopping contained in the interaction.  Since the orbitals are arranged in a
sequence, one can think of the arrangement as a one-dimensional `lattice' system with open or
periodic boundary conditions (sphere and torus respectively).  However, the quartic terms driving
the dynamics are very peculiar from the perspective of lattice models.
Viewed as a lattice (chain), our system is described by a Hamiltonian of the form
\begin{equation}
  H = \sum_{i,k,m} V_{k,m}^{(i)} a_{i+k+m}^\dagger a_{i-m}^\dagger a_{i+k} a_{i} .
\label{eq:H_hopping}
\end{equation}
The hopping terms involve two particles hopping symmetrically outward away from, or symmetrically
inward toward, each other.  This form of the Hamiltonian appears because the system is described in
momentum space, and momentum is conserved.  Since momentum in one direction is coupled to position
in the transverse direction, the process conserves the center of mass position of the pair.
The details of the coupling $V_{k,m}^{(i)}$ depend on the exact geometry, but it has the generic
feature that it falls off at long distances (as a function of either $k$ or $m$) in gaussian fashion
(faster than exponentially), with a length scale scaling as a square root of the number of orbitals
(number of ``lattice sites'').  The coupling is thus neither completely local nor quite non-local.

To highlight the peculiarities of this Hamiltonian, we may compare/contrast with a paradigmatic
single-particle-hopping Hamiltonian that is used in the non-equilibrium literature:
\begin{equation}
  H = - \sum_{i} \Big( a_{i}^\dagger a_{i+1} + a_{i+1}^\dagger a_{i}\Big) + V \sum_{i}
  a_{i}^\dagger a_{i+1}^\dagger a_{i} a_{i+1} . 
\label{eq:H_tightbinding}
\end{equation}
This is a tight-binding model for fermions with nearest-neighbor interactions of strength $V$, and
is equivalent to the $XXZ$ spin chain apart from boundary terms.
Although this is an interacting quartic Hamiltonian, like the Hamiltonian \eqref{eq:H_hopping} to be
examined in this work, the hopping occurs through a quadratic term (single-particle hopping).  This
feature is quite typical of models generally considered in the quantum non-equilibrium literature
\cite{PolkovnikovSilva_RevModPhys2011, Eisert_NatPhys2015, PolkovnikovRigol_AdvPhys2016,
  Mitra_review2018}.  
Clearly, when considered in momentum space, the dynamics in this model will also proceed
through symmetric two particle hopping. However, outside of the FQH context,
we can not interpret this as hopping in real space. We will discuss this further in section \ref{sec:concl}.

We consider dynamics arising from an inhomogeneous initial state.  We pack the $N$ fermions in a
block of $N$ contiguous orbitals, so that there is a block of filled orbitals which spreads into the
rest of the sphere/torus, the rest of the orbitals being initially empty.  With respect to any
subdivision of orbitals, this initial state is a zero-entanglement product state.  This has an
obvious analogy in fermionic tight-binding chains, and is also analogous to a spin chain where some
contiguous block start in a spin-up state and the rest start in a spin-down state, \eg the `domain
wall' state $\ket{\uparrow\uparrow\uparrow\uparrow\downarrow\downarrow\downarrow\downarrow}$.
For tight-binding fermionic chains and spin chains, this type of dynamics by now has been studied in
some detail 
\cite{Antal_PRE1999, KollathSchuetz_PRE2005, Santos_PRE2008, Antal_PRE2008, SantosMitra_PRE2011,
  Haque_PRA2010, MosselCaux_NJP10, Misguich_PRB2013, EislerRacz_PRL2013, VitiStephanHaque_EPL2016,
  HeidrichMeisnerPollmann_PRB2016, BertiniColluraFagotti_PRL2016, EislerEvertz_Scipost2016,
  DubailStephanVitiCalabrese_SciPost2017, JMStephan_JSM2017, MisguichKrapivsky_PRB2017,
  VidmarRigol_PRX2017, ColluraDeLucaViti_PRB2018}.
Related situations, \eg when one/both of the regions are not completely full/empty, are also of
interest and sometimes involve similar physics \cite{HeidrichM-etal_PRA2009, LancasterMitra_PRE2010,
  RibeiroLazaridesHaque_PRA2013, AlbaHeidrichMeisner_PRB2014, PeottaDiVentra_NPhys2015,
  Lancaster_PRE2016, VitiStephanHaque_EPL2016, Pereira_xxz-joining_PRB2017,
  ZnidaricProsen_JPhysA2017, ZnidaricProsen_NatComm2017, Kormos_Scipost2017}.
As the dynamics can be regarded as being initiated by abruptly joining two regions in different
equilibrium states, this process is sometimes known as an `inhomogeneous quench'.  If both parts are
locally ground states of the Hamiltonian, there is a spreading or `transport' of fermions or
magnetization from one region to the other.  For local Hamiltonians, an intermediate
current-carrying region grows linearly as a `light cone', within which a non-equilibrium steady
state emerges.
The form of the non-equilibrium steady state \cite{Antal_PRE1999, Misguich_PRB2013,
  VitiStephanHaque_EPL2016}, the nature of the boundary of the light-cone \cite{EislerRacz_PRL2013,
  VitiStephanHaque_EPL2016}, spreading of entanglement \cite{VitiStephanHaque_EPL2016}, and the
effects of interactions \cite{KollathSchuetz_PRE2005, Santos_PRE2008, Misguich_PRB2013,
  JMStephan_JSM2017} and integrability \cite{BertiniColluraFagotti_PRL2016,
  DubailStephanVitiCalabrese_SciPost2017} are some of the many aspects that have been investigated.
In this literature, the spreading dynamics is driven by single-particle hopping, as opposed to the
correlated pair hopping driving our LLL dynamics.  We expect such `transport' dynamics to be
well-suited for exposing the differences between the two classes of hopping.

A hallmark of isolated quantum systems is that the dynamics can be governed by any part of the
many-body eigenspectrum: the low-energy parts of the spectrum do not necessarily play an important
role.  The reason is that, if the initial state has overlaps with some set of eigenstates, possibly
far from the low-energy part of the spectrum, then (assuming a time-independent Hamiltonian) the
system dynamics does not explore any other part of the spectrum, as there is no tendency toward the
ground state in the absence of external dissipation mechanisms.  Initial states of the inhomogeneous
type we use, typically have largest overlap with eigenstates far removed from the low-energy regime.
The interest in the present Hamiltonian and associated geometries has been largely due to the exotic
topological and fractionalization properties of the ground state and low-lying excited states
\cite{Laughlin_PRL1983, ArovasSchriefferWilczek_PRL1984, MacDonald_arxiv1994, Wen_AdvPhys1995,
  Wen_2003_Book}.  In this work, this class of properties is not expected to play any role, since
the physics explored is not directly related to ground state physics.  Instead of the low-energy
part of the spectrum, we present a study of the \emph{level spacing statistics} of the \emph{full}
many-body spectrum. The level statistics are known to be sensitive to conservation laws and
integrability \cite{BerryTabor_ProcRSoc1977, BohigasEA_PRL1984, Haake_2010_Book,
  Wimberger_2014_Book, RobnikBerry_JPA1986}, which may have nontrivial effects on the
non-equilibrium behaviors.

Although the physics explored here is not directly related to FQH physics, we show results at the
filling at which the best-known FQH state, namely the Laughlin $\nu=1/3$ state
\cite{Laughlin_PRL1983}, appears.  We place $N$ fermions on a torus with $3N$ orbitals or on a
sphere with $3N-2$ orbitals.  The phenomena we describe do not depend strongly on the filling, as
long as the filling is not too close to $0$ or $1$.  Our choice of $1/3$ filling simply reflects the
historical importance of this filling.  The extreme case of unit filling represents an integer
quantum Hall state, which has no dynamics as long as we are confined to a single Landau level.

The precise form of the interaction we use also reflects a FQH tradition: we use the so-called
pseudopotential $V_1$.  This is a short-range potential in real space.  It has the Laughlin
$\nu=1/3$ state as its exact ground state \cite{TrugmanKivelson_PRB1985,
  Haldane_inPrangeGirvinbook_1987, MacDonald_arxiv1994}.  The dynamics we will present depend on the
correlated-pair-hopping nature of the Hamiltonian, but do not depend strongly on the precise form of
the interaction.  When appropriate, we comment on differences or similarities between the
short-range $V_1$ potential and the long-range Coulomb potential, with regards to dynamics and level
statistics.

In Section \ref{sec:geometry}, we provide a description of the geometries and the form of the
Hamiltonian in the orbital basis.  
In Section \ref{sec:initial_state}, we describe our initial states. Considering initial states where
the $N$ fermions are placed on a block of $N$ consecutive orbitals, the position of this block can
still be varied.  On the torus, shifting the block position makes no difference due to translation
symmetry, analogous to a periodic tight-binding chain.  On the sphere, one could consider having the
block starting at one of the poles.  In the chain analogy, this corresponds to the block starting at
one edge of the open chain, so that there is a single `domain wall'; this is strongly analogous to
the literature on inhomogeneous quenches.  We explain however that such a state is inert and has no
dynamics whatsoever, due to conservation of angular momentum. Instead, we primarily focus on the
initial state having the block of fermions centered at the equator.

Section \ref{sec:time_evolution} focuses on real-time dynamics. We present details of the `melting'
process of the equatorial block, such as the intermediate and long-time occupancy/density profiles,
the evolution of entanglement entropies, relaxation timescales etc. We point out differences and
similarities with a `usual' tight-binding model with single-particle hoppings; for example the
evolution of the entanglement entropy profile shows a striking difference.  We examine the `overlap
distribution' --- the overlaps of the initial state with all energy eigenstates. For the torus, the
overlaps are found to be dominated by a few eigenstates at the very top of the many-body spectrum.
As a result, our initial state on the torus has no substantial dynamics, even less so with
increasing sizes. We also examine a shifted initial state on the sphere, which leads to a
non-uniform asymptotic occupancy profile whose overall shape (nearly linear) we are able to explain
using mean-field-like considerations.  

Section \ref{sec:level_statistics} discusses the level spacing statistics.  By identifying
Wigner-Dyson statistics, we demonstrate that the LLL Hamiltonians are not integrable, either with
$V_1$ or with Coulomb interactions.  We demonstrate also that the Hamiltonian on the sphere and in
several momentum sectors on the torus follow GOE (Gaussian orthogonal ensemble) statistics, despite
the system having broken time-reversal symmetry.  We explain this counter-intuitive effect using a
result reported in Ref.\ \cite{RobnikBerry_JPA1986} for single-particle systems, namely, that GOE
statistics may appear in time-reversal-broken systems if there is a unitary symmetry which, combined
with time-reversal, provides an anti-unitary symmetry leaving the Hamiltonian invariant.  To our
knowledge, this is the first time this effect has been observed in a many-body setting.

Concluding comments and discussion are presented in Section \ref{sec:concl}.

\begin{figure*}[tbp]
  \begin{centering}
    \includegraphics[width=.90\linewidth]{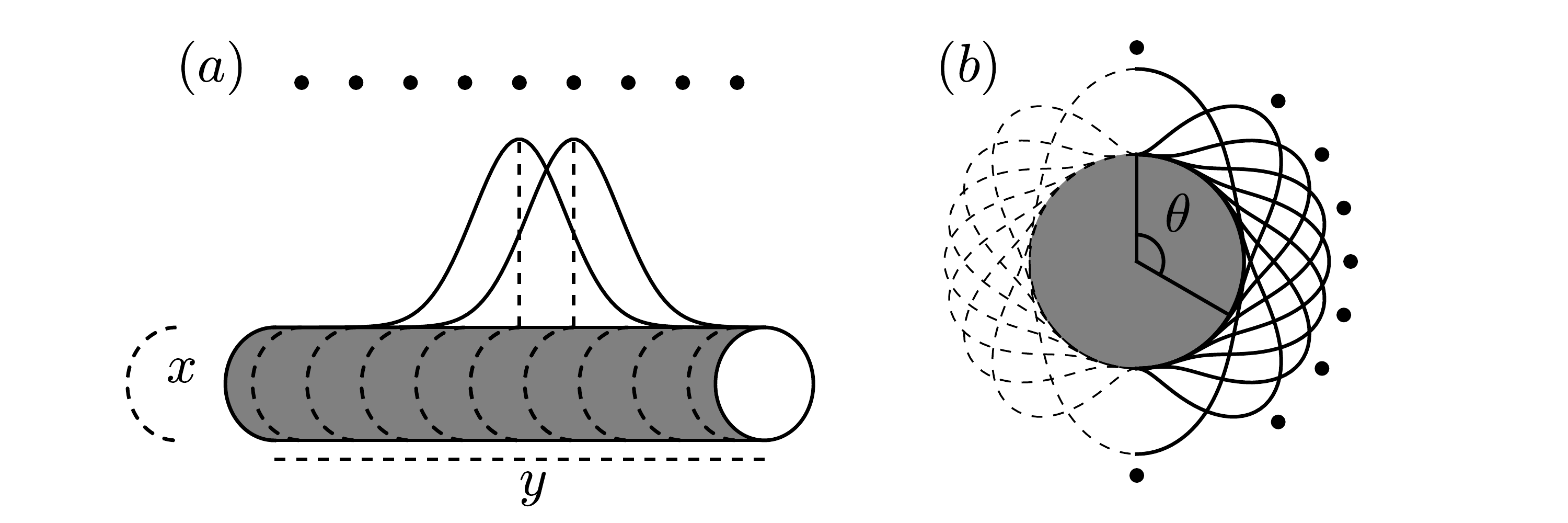}
  \end{centering}
  \caption{Arrangement of LLL orbitals on (a) torus and (b) sphere geometries.  
(a) 
The curves are proportional to the real-space density $|\psi|^2$ for two adjacent orbitals (out of a
total of nine) on a torus.  The dots represent the $y$-position of the density peaks of orbitals.
The torus has its orbitals $\psi_k$ equidistantly  
spaced at positions $y=k L\tau_2/N_\phi$, where $k$ is the momentum quantum number in the $x$
direction.  Each orbital wraps around the torus in the $x$ direction.
(b)
The curves show the real space densities $|\psi|^2$ for all (nine) orbitals on a sphere, each
normalized as $|\psi(\theta)/\psi(\theta_{max})|^2$ for better overall visibility.  
Dots are $\theta$ positions of the density peaks.  The orbitals each wrap around the
sphere along lines of equal latitude.  The orbitals are not spaced equidistantly, but have peaks at
$\theta=\arccos (\frac{m}{Q})$; where $m$ is the orbital index running from $-Q$ to $+Q$.
\label{fig:orbitals} }
\end{figure*}

\section{Geometries and Hamiltonian}\label{sec:geometry}

In this section we introduce the geometries and the forms that the interaction Hamiltonian takes in
the lowest Landau level on the two geometries.
Our goal here is to provide enough details for a reader from outside the field of fractional quantum Hall physics to appreciate the geometry and the forms of the Hamiltonian.

We will first describe the single-particle problem and single-particle eigenfunctions (orbitals) on
both the sphere and torus, followed by discussion of the interacting problem.

\subsection{Single-particle orbitals and quantum numbers}\label{sec:single_particle}

A charged particle (possibly an electron) with mass $m$ and charge $q$, confined to move in two
dimensions ($x$ and $y$), in a magnetic field $B$ pointing in the third ($z$) direction, is subject
to the Hamiltonian $H = \sum_{j=x,y}\frac1{2m}\left(p_j-eA_j\right)^2$.  Here $\vec{A}$ is the
vector potential.  The quadratic Hamiltonian can be diagonalized by introducing ladder operators
$a$, $a^\dagger$ and $b$, $b^\dagger$.  The diagonalized hamiltonian is $H =
\hbar\omega\left(a^\dagger a + \frac 12\right)$, with $\omega= |q|B/m$.  The absence of the $b$
operator signifies an extensive degeneracy with respect to each eigenvalue of $a$.  Each energy band
is called a Landau Level (LL), and the lowest band is the lowest Landau Level (LLL).  We are
interested in the high-field regime such that the separation between Landau levels is far larger than the
interaction energy scale, so that we can focus on the LLL and ignore all other Landau levels.
The single-particle eigenstates within the LLL have a characteristic length scale (`magnetic length')
$\ell_B=\sqrt{\hbar/|q|B}$.

\subsubsection{Torus \label{ref_singlepcle_torus}}
We define the torus as a Bravais lattice with lattice vectors $\mathbf{L_1}$ and $\mathbf{L_2}$, not
necessarily perpendicular to each other.  We define $\mathbf{L_1}=(L,0)$ and $\mathbf{
  L_2}=(\tau_1L,\tau_2L)$, where $\tau=\tau_1+i\tau_2$ is a complex parameter.  This is a standard
parametrization of the torus geometry
\cite{HaldaneRezayi_PRB1985,HermannsSuorsaBergholtzHanssonKarlhede_PRB2008,Read_PRB2009,
  Fremling_JPA2013,FremlingHanssonSuorsa_PRB2014,
  FremlingFulsebakkeMoranSlingerland_PRB2016,Fremling_JPA2016,FremlingMoranSlingerlandSimon_PRB2018}.
Rectangular tori, for which $\mathbf{L_1}$ and $\mathbf{L_2}$ are orthogonal, are obtained for
$\tau_1=0$.  In this case, for $\tau_2\to0$ and $\tau_2\to\infty$, one obtains the `thin torus'
limit, which has been used fruitfully to gain insights into FQH physics \cite{TaoThouless_PRB1983,
  RezayiHaldane_PRB1994, BergholtzKarlhede_PRL2005, SeidelLeeMoore_PRL2005,
  BergholtzKarlhede_JSM2006, SeidelLee_PRL2006, BergholtzKarlhede_PRB2008, SeidelYang_PRB2011,
  OrtizNussinovSeidel_PRB2013, Seidel_PRB2014, Papic_thintorus_PRB2014}.  We will restrict to unit
aspect ratio ($|\mathbf{L_1}|=|\mathbf{L_2}|$ and $|\tau|=1$).  In this case, defining $\tau =
e^{i\theta}$, a square torus is obtained for $\theta=\pi/2$, while a trapezoidal torus with
`hexagonal' symmetry is obtained when the shape is distorted to $\theta=\pi/3$.

While the torus has translational invariance in both lattice directions, the geometry within a LL is
non-commutative, hence the momenta in the two directions cannot be simultaneously specified
\cite{Haldane_PRL1985}.  Only one of the lattice momenta can be used as a quantum number to
distinguish the degenerate single-particle states in the LLL.  The torus area is required to be
$A=2\pi N_\phi \ell_B^2$, where $N_\phi$ is the integer number of magnetic flux quanta that is
piercing the torus.

In the LLL, there are $N_\phi$ linearly independent momentum eigenstates, labeled by the integer
$k$ (single-particle momentum in the $x$ direction).  The eigenstates are expressed exactly as
sums over displaced Gaussian functions\cite{Yoshioka_1983}; however the structure of the orbitals is
more easily visualized from the following approximation:
\begin{equation}
\psi_k(\vec r) \approx \frac{1}{\sqrt{\ell_B
    L\sqrt{\pi}}}e^{-\frac{1}{2\ell_B^{2}}\left(y-\frac{L\tau_2k}{N_\phi}\right)^2}e^{-\i2\pi
  k\left(\frac{x}{L}-\frac{\tau_1}{2N_\phi}k\right)} .
\end{equation}
As shown in \figref{fig:orbitals}(a), these orbitals are extended in the $x$-direction, and
localized in the $y$-direction at $y=-L\tau_2k/N_\phi$ with width $\ell_B$.  Note that position in
the $y$ direction corresponds to momentum in the $x$ direction --- each orbital is localized around
a $y$-position determined by the $x$-momentum quantum number $k$.  This is another feature of the
noncommuntative geometry within LLs.

One could of course have chosen to work with eigenstates of the momentum in the $y$ direction; in
that case the orbitals would wrap around the torus in $y$ direction and be localized in the $x$
direction around a $x$-position set by the $y$-momentum quantum number.

\subsubsection{Sphere}

The so-called Haldane sphere \cite{Haldane_PRL1983, HaldaneRezayi_PRL1985,
  Haldane_inPrangeGirvinbook_1987, FanoOrtolaniColombo_PRB1986, HaldaneRezayi_PRL1988,
  Jain_2007_Book} has a magnetic monopole in its center, such that it is pierced by $N_{\phi}=2Q$
flux quanta, corresponding to a magnetic field $\mathbf{B}=\frac{2Q\phi_0}{4\pi R^2}\hat r$.  Here
$2Q\phi_0$ is the flux through the spherical surface.  The Dirac quantization condition dictates
that $N_\phi$ is an integer.

The Hamiltonian of a single particle on the sphere may be written as 
$H = \frac{\hbar^2}{2m}\left|\hat{\bf R}\times (-\i\mathbf{\nabla} - \frac{q}{\hbar c}\mathbf{A}(\mathbf{
\Omega}))\right|^2$.
We use the notation $(\phi,\theta)$ for azimuthal and polar angles.  The vector potential $\mathbf{
  A}=-\frac{\hbar cQ\phi_0}{eR}\cot \theta\hat\phi$ has two Dirac strings, one through the north
pole and one through the south pole.  Unlike the torus, there is no momentum quantum number,
however, a magnetic analogue of angular momentum ($\mathbf{L}$) can be defined
\cite{Haldane_PRL1983,FanoOrtolaniColombo_PRB1986, Jain_2007_Book}.
The Hamiltonian can be rewritten as $H = L^2 -Q^2$.  Both $L^2$ and $L_z=-\i\partial_\phi$ are
operators commuting with the Hamiltonian.  The LLL orbitals (single-particle eigenfunctions) can be
labeled by the $L_z$ quantum numbers $m$.  The orbitals labeled in this way wrap around the sphere
as lines of latitude.  There are $2Q+1=N_\phi+1$ orbitals, $m=-Q,-Q+1,\ldots,Q-1,Q$, running from
the south pole to the north pole.
Just as the $x$ momentum on the torus is coupled to the $y$ position of the
orbitals, on the sphere the $z$-component of angular momentum is coupled to the polar position of
the orbitals.  The eigenfunctions have the form 
\begin{equation}
\psi_m \propto e^{\i m\phi}[\sin(\theta/2)]^{Q-m}[\cos(\theta/2)]^{Q+m}
\end{equation}
and are localized around $\theta=2\arctan(\sqrt{\frac{Q+m}{Q-m}}) =\arccos(m/Q)$.  A cartoon of the
arrangement of orbitals is shown in \figref{fig:orbitals}(b).

\begin{figure*}[tbp]
\begin{center}
    \includegraphics[width=.90\linewidth]{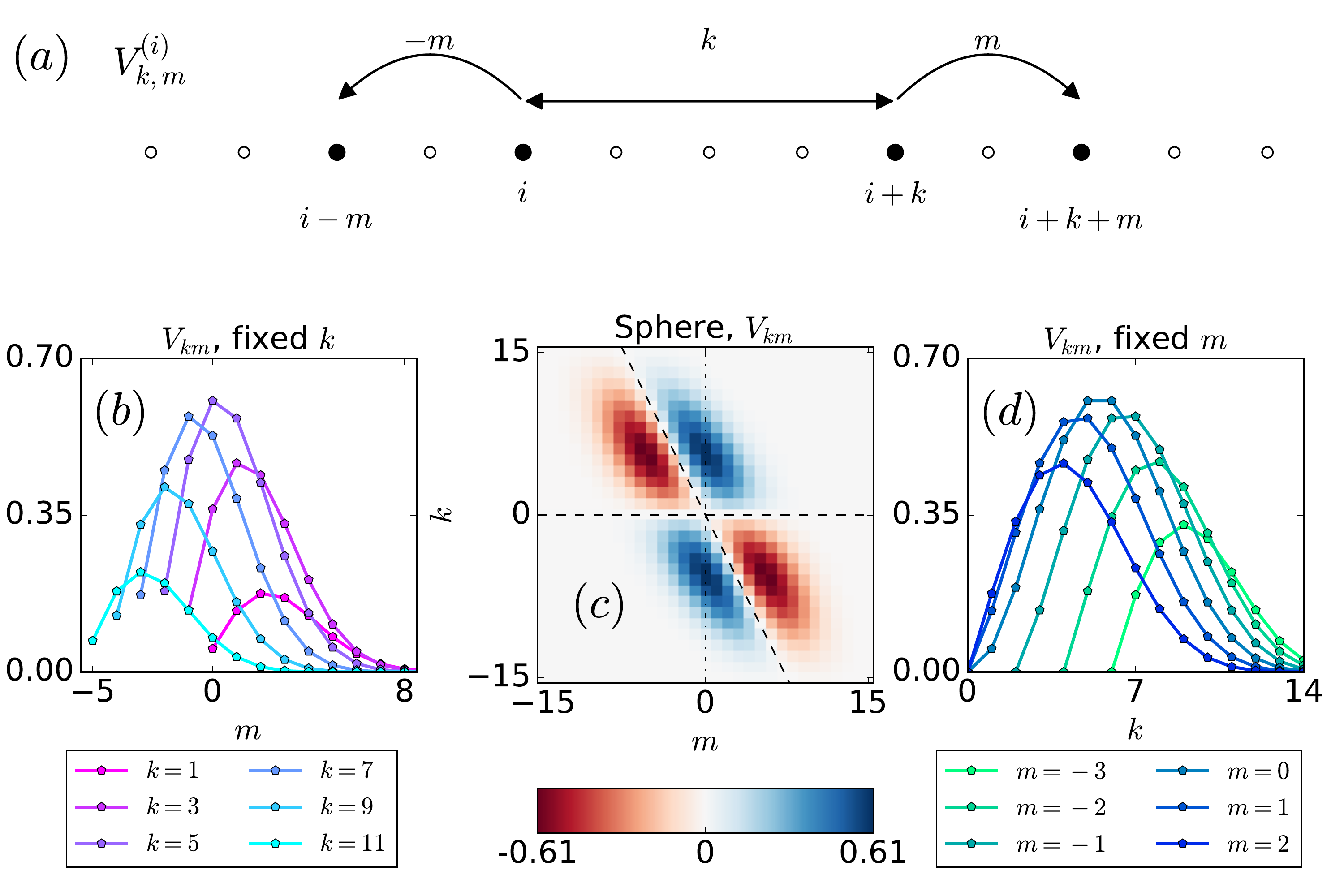}
\end{center}
\caption{    
(a)
Illustration of the correlated symmetric hopping procedure that governs the dynamics. The hopping
element $V^{(i)}_{k,m}$ causes the electrons at site $i$ and site $i+k$, to hop $m$ orbitals away
from each other, making the new distance $k+2m$.  On the torus, due to translation invariance
$V^{(i)}_{k,m}=V_{k,m}$ for all $i$.
Due to electron indistinguishably, 
$V^{(i)}_{k,m}=-V^{(i)}_{k,-k-m}=-V^{(i+k)}_{-k,k+m}=V^{(i+k)}_{-k,-m}$. 
(b)
Values of $V_{km}$ for fixed some fixed $k$. Data points for $m<-k/2$ have been excluded as they are redundant.
(c)
Values of $V_{km}$ in the $k$-$m$ plane.
(d)
Values of $V_{km}$ for fixed some fixed $m$. Data points for $k<-2m$ and $k\leq0$ have been excluded
as they are redundant.
Panels (b)-(d) all correspond to a sphere with fixed $i=15$ for $2Q=30$
  \label{fig:hopping}
}
\end{figure*}

\subsection{Many interacting fermions}

We now consider $N$ interacting fermions.  In the original FQH literature, the Coulomb interaction
was most natural to consider as it is the physical interaction between electrons.  Another
interaction that also has a long history is the short-range interaction that produces the Laughlin
state \cite{Laughlin_PRL1983} as its exact ground state --- this is the Trugman-Kivelson interaction
$\nabla^2\delta(\vec{r})$ \cite{TrugmanKivelson_PRB1985} projected onto the LLL.  This interaction
penalizes two particles interacting via the $p$-wave channel (the $s$-wave channel being anyway
forbidden by fermionic antisymmetry).  It is thus also an interaction form that could conceivably be
realized in cold-atom experiments with spin-polarized fermions in reduced dimensions
\cite{KoehlEsslinger_PRL2005}.
When projected onto the LLL, two-body interactions are conveniently described in terms of
pseudopotentials $V_{m}$ \cite{Haldane_inPrangeGirvinbook_1987, Haldane_PRL1983,
  RezayiHaldane_PRB1994, MacDonald_arxiv1994, Jain_2007_Book}.  The Trugman-Kivelson interaction
leads exactly to the $V_1$ pseudopotential; hence this is also known as the $V_1$ interaction.
Given its historical importance and somewhat easier form, we will focus on the $V_1$ interaction in
this work, including occasional comparisons with the Coulomb interaction.  

To treat a central isotropic interaction potential $V(|\vec{r}-\vec{r'}|)$, on the torus one needs
to use a periodicized form, $\tilde{V}(\vec r)=\tilde{V}(\vec r + n\mathbf{L_1}+m\mathbf{L_2})$, while on
the sphere the arc distance between $\vec{r}$ and $\vec{r'}$ is used.  In the basis of orbitals,
such an interaction (either Coulomb or $V_1$) takes the form $H = \sum_{i,k,m} V_{k,m}^{(i)}
a_{i+k+m}^{\dagger}a_{i-m}^{\dagger}a_{i+k}a_{i}$ declared in the Introduction,
Eq.\ \eqref{eq:H_hopping}.  
This Hamiltonian includes both static repulsion (for $m=0$) and long-range correlated pair hopping
terms ($m\neq0$).

On the torus, the interaction is translationally invariant, \ie $V_{k,m}^{(i)}$ is independent of
$i$.  For a long torus, it takes the form $V_{k,m} \propto (k^2-m^2)e^{-(2\pi\ell_B/L)^2(k^2+m^2)}$
\cite{RezayiHaldane_PRB1994}.  The amplitude of correlated hopping falls off rapidly with $k$ and
$m$ but nevertheless remains finite at long ranges, which allows the dynamics to get initiated
starting from our block initial states.  The length scale of the Gaussian in units of $\ell_B$ is
set by $L$, the circumference in the transverse direction.  For tori of unit aspect ratio, the
interaction length scale thus scales as $\sim\sqrt{N_\phi}$.  Viewed as an interaction on a lattice
with $\sim N_{\phi}$ sites, $V_{k,m}^{(i)}$ cannot be viewed either as short-range or as long-range
in any conventional meaning of these terms.

Thus, although the Trugman-Kivelson interaction is short range in real space, when the system is
viewed as a tight-binding chain of orbitals, it leads to fermions with correlated hoppings that are
not very short-range, as shown in \figref{fig:hopping}(a).

On the sphere, the interaction $V_{k,m}^{(i)}$ depends on $i$, reflecting the lack of translation
invariance in this geometry.  The dependence on $k$ and $m$ is more complicated to write than is the
case on the torus, but has similar qualitative form, and is shown in \figref{fig:hopping}(b-d).  In
particular, we expect the length scale of the interaction to scale on average as
$\sim\sqrt{N_\phi}$.

\subsubsection{Symmetry Sectors}\label{sec:symmetry_sectors}

On the sphere, there exists a magnetic analogue of angular momentum ${\bf L}$, that preserves all
the commutation relations of usual angular momentum operators.  We may thus choose to use the $L_z$
operators to enumerate the basis of single particle orbitals, and diagonalize the Hamiltonian within
this basis.  In the many-body system, there now exists a total $L_z$ and corresponding total $L^2$,
which both commute with the Hamiltonian. 
Block diagonalizing the Hamiltonian with respect to $L_z$ is trivial, but
dividing the Hilbert space into $L^2$ sectors is tricky and most often not implemented directly in
numerical calculations. Our calculations are performed in fixed total $L_z$ sectors of the Hilbert
space.  Our time evolution is performed with initial states that are not eigenstates of total $L^2$,
thus multiple $L^2$ sectors are involved.  When separation of $L^2$ sectors is needed (in the level
statistics analysis of Section \ref{sec:level_statistics}), we add a term $\alpha L^2$ to the
hamiltonian and post-process the data by sorting the eigenvectors by $\left<L^2\right>$.

On the torus there are two generators $t_1$ and $t_2$ which measure the (magnetic) momentum ($k_1$
and $k_2$) of the single particle orbitals in the two lattice directions.  These two operators do
not commute, but satisfy $t_1t_2=e^{i\frac{2\pi}{N_\phi}}t_2t_1$, which shows that the maximal sets
of commuting operators are $t_1,t_2^{N_\phi}$ or $t_1^{N_\phi},t_2$.  For the many-body state we may
define the operators $T_1=\prod_{j=1}^N t_1^{(j)}$, $T_2=\prod_{j=1}^N t_2^{(j)}$, which measure the
total momentum $K_1$ and $K_2$ of the many body state.  If the filling fraction is
$\nu=N/N_\phi=1/q$, then $T_1T_2=e^{i\frac 1q}T_2T_1$.  This shows that the maximal sets of
commuting operators are $T_1,T_2^q$ or $T_1^q,T_2$.  As a consequence, some combinations of $K_1$
and $K_2$ are redundant.  Choosing \eg $T_1,T_2^q$, means that $K_1=0,\ldots,N_\phi-1$, whereas
$K_2$ and $K_2+N$ give the same $T_2^q$ eigenvalue, reducing $K_2$ to $K_2=0,\ldots,N-1$.  Finally,
since $(K_1,K_2)$ and $(K_1+N,K_2)$ have the same energy spectrum, we may define a reduced many-body
Brillouin zone using $(K_1,K_2)$, where $K_1,K_2=0,\ldots,N-1$, as in previous work
\cite{Haldane_PRL1985, BernevigRegnault_PRB2012, PapicHaldaneRezayi_PRL2012,
  RepellinMongSenthilRegnault_PRB2017, FremlingMoranSlingerlandSimon_PRB2018}.

\section{Initial state}\label{sec:initial_state}

The Hamiltonian is number-conserving, and the number of fermions, $N$, is thus fixed during
dynamics.  In considering scaling with system size, we keep the filling fixed.  Since our
Hamiltonian and geometries are motivated by the literature on fractional quantum Hall states, we use
the filling corresponding to the most prominent such state, namely the Laughlin state at filling
one-third.  For the torus, this means that we have $N$ fermions in exactly $3N$ orbitals.  For the
sphere geometry, the Laughlin state appears when $N$ fermions are placed on a sphere with $3N-2$
orbitals.  Here $2$ is the so-called `shift' and is present due to the curvature of the sphere
\cite{Haldane_PRL1983, HaldaneRezayi_PRL1985, Haldane_inPrangeGirvinbook_1987,
  FanoOrtolaniColombo_PRB1986, Jain_2007_Book, HaldaneRezayi_PRL1988}.

Our choice of filling means that the ground state is a topological (Laughlin) state.  However, since
we are looking at non-dissipative dynamics, the ground state or low-energy segment plays no
distinguished role; the eigenstates relevant to our dynamics are far from this part of the spectrum.
In fact, similar dynamics is expected for slightly altered filling fractions, even though such a
sector would have utterly different properties from the point of view of studying ground-state
properties.

We consider initial states which are product states in the orbital basis: the $N$ fermions are
placed in $N$ orbitals.  We will focus on initial filling of $N$ \emph{successive} orbitals:
\begin{equation}  \label{eq:initial_block_state} 
\ket{\psi(0)} ~=~ a^{\dagger}_{j+1} a^{\dagger}_{j+2}\ldots a^{\dagger}_{j+N}
\ket{\mathrm{vacuum}} 
\end{equation}
This is the analog of fermionic dynamics on a tight-binding lattice where all the fermions are
initially packed on adjacent sites 
\cite{EislerRacz_PRL2013, VitiStephanHaque_EPL2016, HeidrichMeisnerPollmann_PRB2016,
  DubailStephanVitiCalabrese_SciPost2017},
or spin-$\frac{1}{2}$ dynamics starting from a configuration in which all the up-spins are on
adjacent sites, \ie starting from a domain-wall state
\cite{Antal_PRE1999, KollathSchuetz_PRE2005, Santos_PRE2008, Antal_PRE2008, SantosMitra_PRE2011,
  Haque_PRA2010, MosselCaux_NJP10, Misguich_PRB2013, BertiniColluraFagotti_PRL2016,
  EislerEvertz_Scipost2016, JMStephan_JSM2017, MisguichKrapivsky_PRB2017,
  ColluraDeLucaViti_PRB2018}.

On the torus, the starting point ($j$ in Eq.\ \eqref{eq:initial_block_state}) of the block of
fermions does not matter; translation symmetry guarantees that the dynamics is equivalent for any
starting position.  For the sphere, however, the positioning of the string affects the dynamics
drastically.

For the sphere, we could consider placing the fermions in the $N$ orbitals closest to one of the
`poles'.  In the `tight-binding chain' picture, this would mean unit filling for the first third or
last third of the chain, and zero filling for the rest of the chain.  This would then be analogous
to the dynamics considered in, \eg Refs.\ \cite{Antal_PRE1999, KollathSchuetz_PRE2005,
  Santos_PRE2008, Antal_PRE2008, SantosMitra_PRE2011, Haque_PRA2010, MosselCaux_NJP10,
  Misguich_PRB2013, EislerRacz_PRL2013, VitiStephanHaque_EPL2016, HeidrichMeisnerPollmann_PRB2016,
  BertiniColluraFagotti_PRL2016, EislerEvertz_Scipost2016, DubailStephanVitiCalabrese_SciPost2017,
  JMStephan_JSM2017,
  MisguichKrapivsky_PRB2017, VidmarRigol_PRX2017, ColluraDeLucaViti_PRB2018}, for spin or
tight-binding systems, when the initial state has all fermions or all $\uparrow$-spins on one side
of the chain.  In such cases, the single-particle hopping leads to transport of particles from the
filled side to the empty side, with a current-carrying non-equilibrium steady state arising
dynamically within the light cone region.
In our system, however, this type of initial state leads to an inert state with no dynamics
whatsoever.

The lack of dynamics starting from a polar-cap initial state on the sphere can be understood in
several ways.  The sphere Hamiltonian conserves the $L_z$ quantum number (introduced in
Sec. \ref{sec:geometry}), which is maximal for this state.  That means that this state belongs to a
symmetry sector of the Hilbert space composed of a \emph{single} state, namely the polar-cap state.
A 1-dimensional Hilbert space, of course, has no dynamics.  Alternatively, one can consider the
correlated pair hopping process, which can occur only if two particles can hop symmetrically
outwards or inwards,  \figref{fig:hopping}(a).  When the particles are packed on the first $N$
orbitals starting from the boundary, only hoppings in one direction are possible, hence there is no
configuration space available for pair hopping, resulting in a frozen state.  (The situation can be
described using terrestrial analogy: `The polar-cap state is frozen.') 

Therefore, we consider dynamics where the block of $N$ fermions are near the equator.  In fact we
will mostly concentrate on the initial state covering a symmetric band around the equator.  However,
in Section \ref{sec:Lopsided} we also show some interesting dynamical effects for `lopsided' initial
states.

\begin{figure}[tbp]
  \begin{center}
    \includegraphics[width=0.45\linewidth]{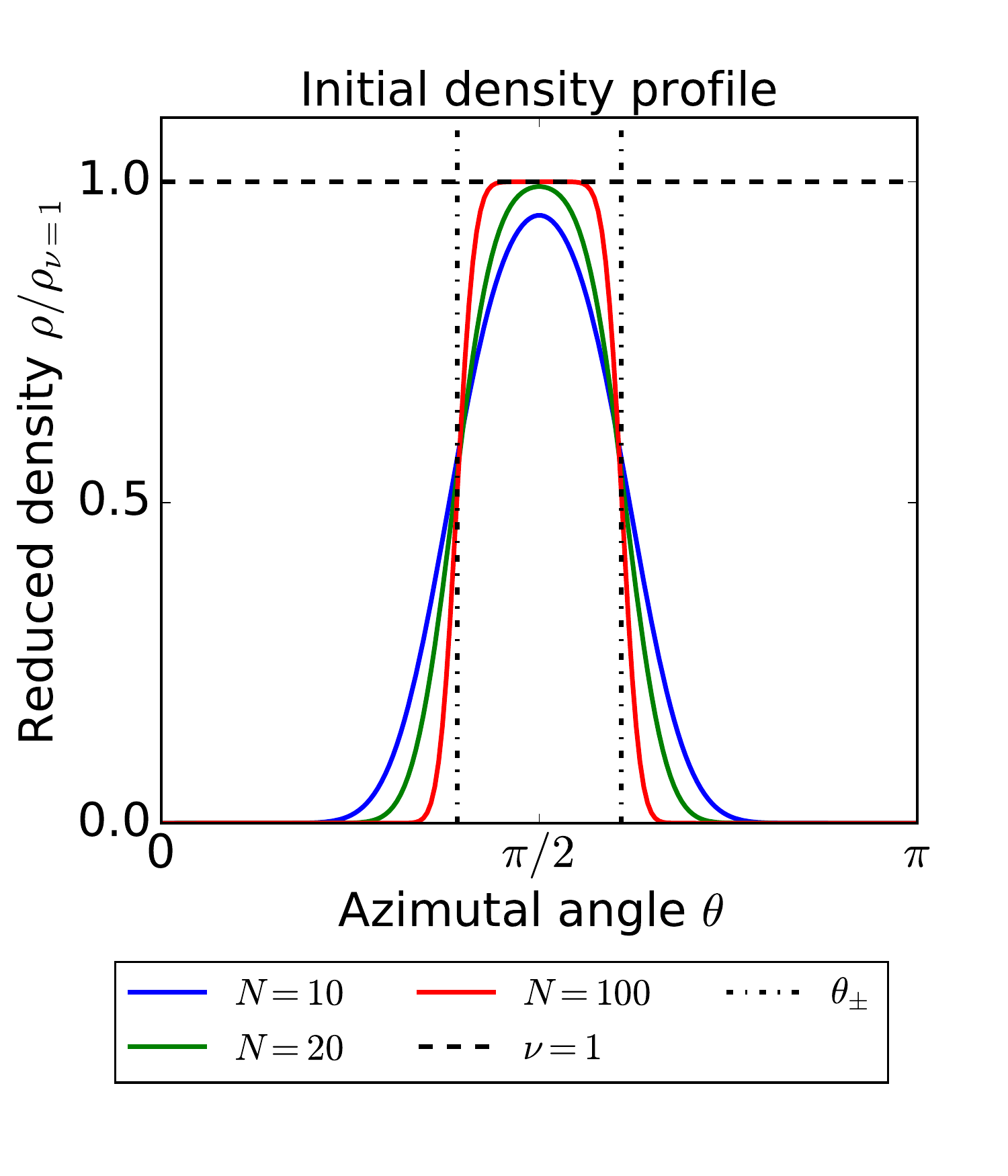}
  \end{center}
  \caption{ The real space density profile $\rho(\theta)$ of the equatorial initial state on the
    sphere, for various system sizes.  To make the comparison of different $N$ possible, the density
    $\rho(\theta)$ is re-scaled by the corresponding density
    $\rho_{\nu=1}=\frac{2Q+1}{4{\pi}R^2}=\frac{3N-2}{4{\pi}R^2}$ that the filled LLL ($\nu=1$ )
    would have.  For larger systems a well defined density plateau emerges near the equator over the
    angles $\theta_-<\theta<\theta_+$, where
    $\theta_{\pm}\approx2\arccos(\mp \nu)=\arccos(\mp \frac{1}{3})$.
    The density decreases from $\rho=\rho_{\nu=1}$ to $\rho=0$ over a distance of the order of the
    magnetic length $\ell_B$.  For system sizes numerically treated in this work ($N\approx10$),
    such a density plateau has not yet formed.}
  \label{fig:initial_config_realspace}  
\end{figure}

While the orbital description is very convenient, one might also want to know what our initial
states look like in real space.  Since the orbitals are loosely localized in space, one expects
that, at large $N$, a region of filled orbitals corresponds to a region with a nearly constant real
space density equal to the maximal density possible in the lowest Landau level. This density is the
real space density observed in the integer quantum Hall state with a completely filled lowest Landau
level, $\nu=1$.  \figref{fig:initial_config_realspace} shows the real space density for various
particle numbers. While this correspondence is correct for large $N$, there are significant
deviations for $N\sim10$, \ie for system sizes accessible numerically.  The initial real space
density profile then has a bell-shaped form.

In \figref{fig:initial_config_realspace}, for large enough $N$, the particles in the initial state
fill up a region near the equator which covers less than one-third of the azimuthal angle.  The
reason is that orbitals in the equatorial region (which are longer) have smaller width than orbitals
in the polar region (which are shorter), because each orbital must cover the same area.  If there is
a plateau, the real-space density falls off at both edges of the plateau with a fall-off width of
the order of a magnetic length $\ell_B$; the decay is faster than exponential.

Although the large-$N$ behavior has a plateau structure, for the dynamical effects  presented in
this paper, it is simply more appropriate to think of the initial state having a gaussian-like
real-space density profile, corresponding to the numerically accessible $N$ values.

\section{Time evolution}\label{sec:time_evolution}

In this section, we present our results on the time evolution of initial states of the type
\eqref{eq:initial_block_state}.  Most of the analysis is for the equatorial initial state on the
sphere. 

We first describe the evolution of the orbital occupancies (\ref{sec:OrbEvol}), considering
snapshots of evolving orbital occupancy profiles and the deviation from the final uniform state.
Subsection \ref{sec:realspace_densities} describes the same evolution in terms of the real-space
densities.  We then present the evolution of the entanglement entropy (\ref{sec:entanglement}), in
particular a contrast with the way entanglement entropy grows in local models with single-particle
hopping.  These three sections focus on the equatorial initial state. 

Subsection \ref{sec:Lopsided} considers a `lopsided' initial state, displaced from the equator.  The
correlated-pair-hopping nature of the Hamiltonian (or equivalently, $L_z$ conservation) now
manifests itself in a remarkable manner --- the long-time state is also lopsided, and for a range of
displacements has a remarkably linear form.  We explain the near-linear shape using mean-field
arguments for the pair-hopping Hamiltonian.

The section ends with a presentation of the overlaps of the equatorial initial state with all
eigenstates of the Hamiltonian, the so-called `overlap distribution' (\ref{sec_overlapdist}).  This
is compared with a tight-binding model and the torus case.  In the torus case, the form of the
overlap distribution leads to the conclusion that the block initial state on the torus is frozen for
larger sizes; we confirm this by showing orbital occupancy dynamics for a sequence of sizes.

The numerical results presented in this work are all obtained by exact numerical diagonalization.
To present the numerical data, we have used the gap $\Delta$ in the ground state sector (the
so-called `neutral gap') as the unit of energy.  This is the energy difference between the Laughlin
(ground) state and the first excited state in the same $L_z$ sector or the same $(K_1,K_2)$ sector.
This is a well-studied energy scale and is of the same order of magnitude as the larger interaction
terms (static or pair-hopping); hence we regard it as a reasonable analogue of the usual practice in
the non-equilibrium literature of using the hopping term to define the units of energy.  Time is
measured in units of $\hbar/\Delta=1/\Delta$.

\subsection{Evolution of orbital occupancies}\label{sec:OrbEvol}

\begin{figure*}[tb]
  \includegraphics[width=0.90\linewidth]{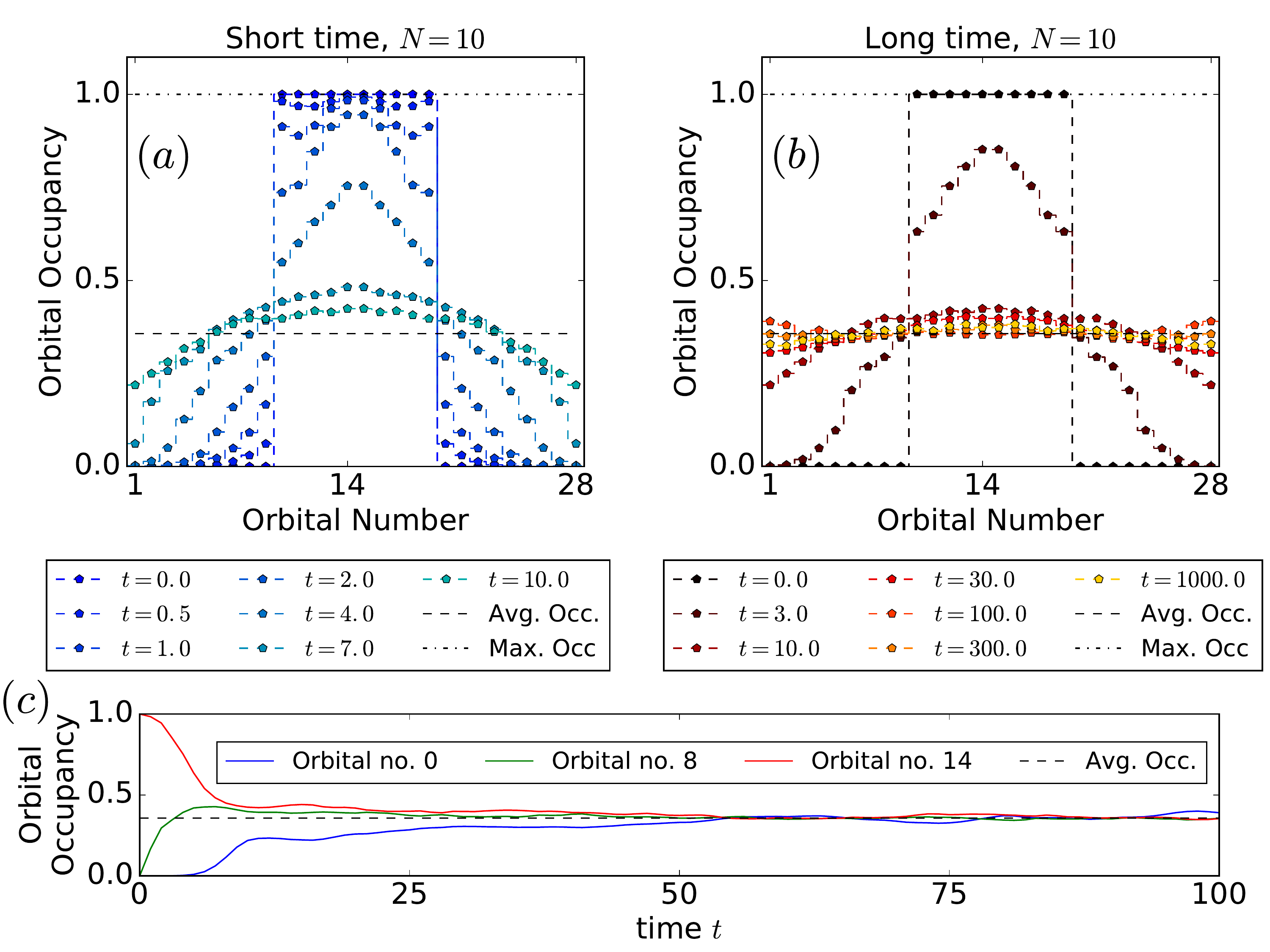}
\caption{Time evolution of the orbital occupancies starting from the equatorial initial state,
  $N=10$ fermions on a sphere.  (a,b) Snapshots of the full occupancy distribution.  At $t=0$ the
  state is a product state, occupying only the central orbitals.  At $t\sim8$ (a, magenta) the
  discontinuity in the distribution has vanished and the occupancy is a ``smooth'' function of
  orbital position.  At $t\gtrsim30$ (b, cyan) the occupancy is near-homogeneous apart from small
  fluctuations.
(c) Occupancies of three individual orbitals are shown as function of time --- the polar orbital, an
  orbital near the initial block edge, and an equatorial orbital. The polar orbital occupancy
  has the slowest relaxation.  
\label{fig:orb_dens_evol}
}
\end{figure*}

In \figref{fig:orb_dens_evol}(a,b), the dynamics starting from the equatorial initial state on the
sphere is shown through successive snapshots of the orbital occupancies.  The two panels show the
same dynamics at shorter and at longer timescales.  The fermions gradually spread out from their
initial position and occupy all the orbitals.  In the very long-time limit, the occupancies approach
a `flat' distribution in which each orbital is equally occupied; this is shown also through
individual orbital occupancy evolutions in \figref{fig:orb_dens_evol}(c).  Of course, for any finite
size, the occupancies cannot relax completely and at long times fluctuate around the average
long-time distribution.

It is tempting to compare the melting at either edge to `domain wall' melting in inhomogeneous
quenches driven by a regular tight-binding Hamiltonian with nearest-neighbor single-particle hopping
\cite{Antal_PRE1999, Misguich_PRB2013, EislerRacz_PRL2013, VitiStephanHaque_EPL2016}.  Detailed
similarity is perhaps not expected, because in our case the melting of the two domain walls are
correlated and not independent local processes.  In the dynamics here, the two discontinuities in
orbital occupancy survive for quite a while, until $t\sim8$, \figref{fig:orb_dens_evol}(a), and it
is only after this time that a smooth profile is obtained.  (With nearest-neighbor single-particle
hopping, the discontinuity generally smoothens out rapidly.)  Our interaction is not strictly
short-range and hence a clean light cone is not expected.  Nevertheless, the spreading out is
gradual --- the empty orbitals near the band get occupied first, and then the occupancy spreads
gradually toward the poles.  This presumably reflects the fact that the $V_{km}$ matrix elements
generally fall off with $m$ (\figref{fig:hopping}).
As a clear light cone is not expected, it is unclear whether or not at larger sizes a spatially
distinguishable region of current-carrying `nonequilibrium steady state' \cite{Antal_PRE1999,
  Misguich_PRB2013, EislerRacz_PRL2013, VitiStephanHaque_EPL2016} emerges.  For the sizes accessible
in our calculations, this question cannot be meaningfully addressed.

One effect of our peculiar (correlated-hopping) interaction is visible at very short times
($t\lesssim1$) --- the orbital occupancy profile is non-monotonic at these times.  Since $V_{km}$
falls off with both $k$ (except for very small values of $k$) and $m$, hopping of pairs from farther
inside the band might be more favorable than hopping from the very edges.  As the two effects ---
dependence on $k$ and dependence on $m$ --- compete, one can obtain a non-monotonic effect; this
shows up in the short-time profile. At larger times, the effect is washed out.  This mechanism for
non-monotonicity is described in more detail in \ref{app:non-monotonic}.

Another contrast with spreading dynamics driven by single-particle hopping is that, in our case, we
do not observe any reflection after the fermionic liquid reaches the poles (edges).  In propagation
dynamics on regular tight-binding chains, there is almost always reflection, even in the presence of
strong interactions.

We now turn to the long-term dynamics, \figref{fig:orb_dens_evol}(b,c).  The occupancies approach a
uniform distribution at long times.  The occupancy at any orbital approaches the average occupancy
value $N/(3N-2)$ and then oscillates around this value, \figref{fig:orb_dens_evol}(c).  The
relaxation to the average occupancy value is faster in the region near the boundary of the initial
state, away from both equator and poles.  The orbitals near the pole take a noticeably long time to
settle to the final value, \figref{fig:orb_dens_evol}(c).  This may be regarded as an effect of the
correlated-hopping nature of the Hamiltonian.  Those $V_{km}$ processes which are symmetric around
the equator would need to have a large $k$ or $m$ if they were to populate the polar-region
orbitals, and are hence suppressed.  For non-symmetric $V_{km}$ processes with smaller $k$ and $m$,
the partner that also hops would have to hop from a smaller density to a larger density region; such
processes are thus also not very effective.  

The latter effect can be formulated intuitively in `hydrodynamic' language as follows.  For a
single-particle-hopping Hamiltonian, density gradients drive particle flow, as hoppings from a
high-density to a low-density site is kinematically favored.  On the other hand, a
correlated-pair-hopping process involves two particles hopping away from (or in toward) a central
region; this will be favored if the densities are lower (resp. higher) on both sides of the central
region. Thus correlated hopping leads to flow driven by second derivatives of density. This
intuitively explains why peaked structures like the short-time density profile lead to rapid
spreading out, but the final density inhomogeneity near the ends of the `chain' are not so easily
corrected by the type of hopping we have.  Later in Section \ref{sec:Lopsided}, we will explicitly
show a situation where a density gradient leads to no flow.

\begin{figure*}[tbp]
  \includegraphics[width=0.90\linewidth]{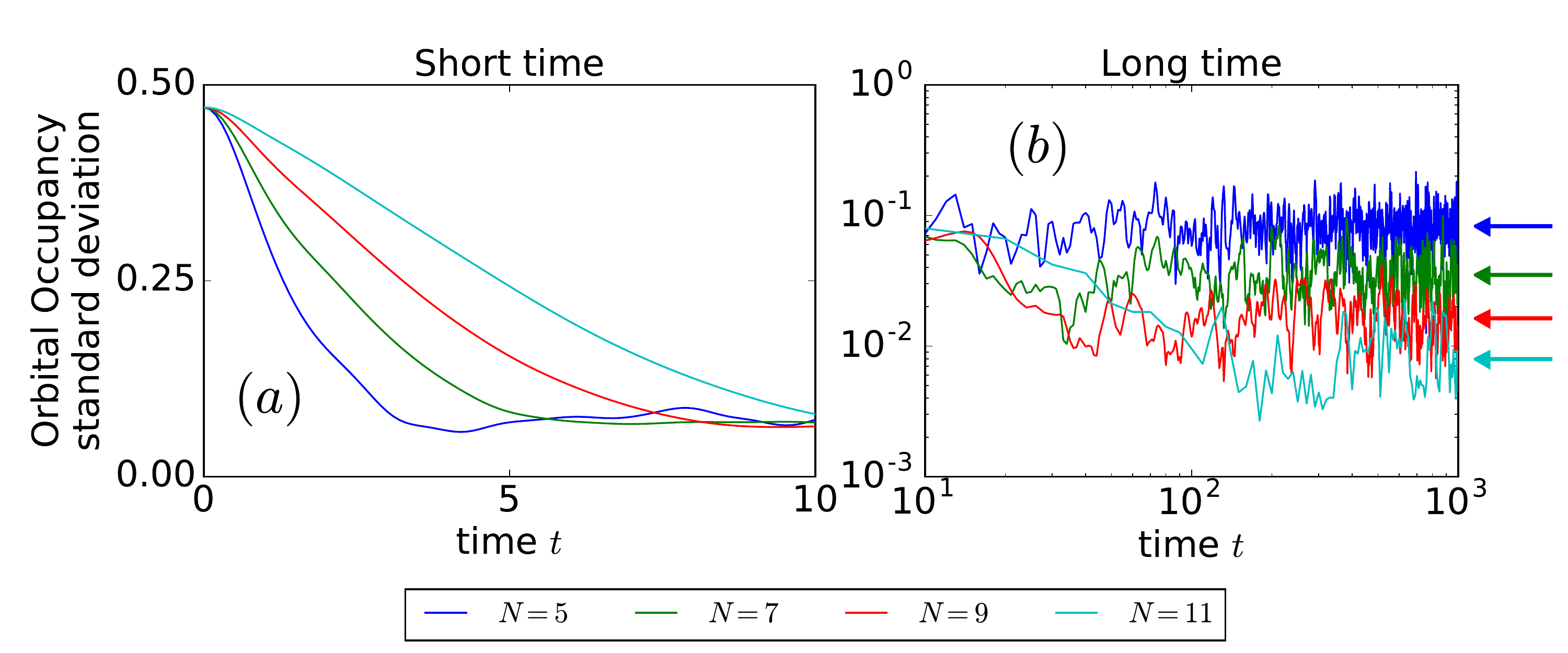}
\caption{Time evolution of the standard deviation $\sigma$ of orbital occupancies, starting from the
  equatorial initial state on a sphere.  The two panels show the same processes for shorter and
  longer times.
At $t=0$, $\sigma=\sqrt{\bar{n}-\bar{n}^2}$, with $\bar{n}=N/(3N-2)$. To make the comparison cleaner
we have therefore re-scaled the standard deviation by $\chi=\sqrt{\frac{ \nu -\nu^2}{\bar
    n-\bar{n}^2}}$, where $\nu=1/3$; each curve starts at the same value.
(a) A first relaxation occurs over time scale $t\sim N$ required for the quantum fluid to reach the
north and south pole of the sphere.  (b) After that there is a slower relaxation period that goes on
for a time scale which grows seemingly super-linearly with $N$.  Finally there is the finite-size
fluctuation around the long-time (geometrical) average value, indicated by the colored arrows.
\label{fig:variance}
}
\end{figure*}

The approach to the final distribution can be visualized by plotting the deviation from the uniform
distribution. This deviation is quantified, for example, by the standard deviation $\sigma$ of the
orbital occupancies.  The time evolution of this quantity is shown for several sizes in
\figref{fig:variance}.  
The initial fast decrease of the variance, \figref{fig:variance}(a), is roughly proportional to the
system size, \ie proportional to the number of orbitals the fermions must spread into.  This is
similar to what one would expect for a short-range hopping, in which case there is a light-cone
effect of information/particles spreading out at a linear rate.  We show here that the same feature
is visible for our long-range correlated hopping; presumably this aspect is not very different from
short-range models because of the rapid decay of matrix elements $V_{km}$ with large $k$ or $m$.  In
\figref{fig:variance}(b), there appears to be a second slower relaxation period, which lasts until
some timescale increasing rapidly (possibly exponentially) with system size; unfortunately the
available sizes do not allow a clean investigation of this time scale.

At longer times the orbital occupancies each fluctuate around the average value, which is seen in
\figref{fig:variance}(b) as the final fluctuating behavior of the standard deviation $\sigma$.  The
strength of the fluctuation decreases with system size.  The data is consistent with an exponential
decrease with the system size --- on the logarithmic plot scale, the average final value of $\sigma$
(shown using left-pointing arrows) decreases in roughly equal steps as $N$ is increased.  The
long-time value of the standard deviation is an average measure of the fluctuation strength of the
individual orbital occupancies.  In the recent literature on the non-equilibrium dynamics of local
observables, the long-time fluctuations have been studied in various systems
\cite{RigolOlshanii_Nature2008, Reimann_PRL2008, GramschRigol_PRA2012, ZiraldoSilvaSantoro_PRL2012,
  VenutiZanardi_PRE2013, HeSantosRigol_PRA2013, PastawskiSantos_PRE2013, ZiraldoSantoro_PRB2013,
  VenutiZanardi_PRE2014, KiendlMarquardt_PRL2017, WangTong_Fibonaccidynamics_JSM2017}.  The
magnitude of fluctuations around the long-time average value is understood to generically decrease
exponentially with system size.  This fluctuation strength is determined by the average magnitude of
the off-diagonal matrix elements of local observables between eigenstates of the Hamiltonian
\cite{RigolOlshanii_Nature2008, Rigol_PRA2009, SantosRigol_PRE2010, KhatamiSrednickiRigol_PRL2013,
  SteinigewegPrelovsek_PRE2013, BeugelingHaque_PRE2015, MondainiRigol_PRE2017,
  McClartyHaque_arxiv2017}.  At least for non-integrable Hamiltonians, this decreases exponentially
with system size (or more precisely, as $\mathcal{D}^{-1/2}$, where $\mathcal{D}$ is the Hilbert
space dimension), leading to an exponential decrease of the long-time fluctuation, in our case
measured by the long-time average of $\sigma$.  The size-dependence of the long-time fluctuations
in our system, \figref{fig:variance}(b), is thus similar to the generically expected behavior for
non-integrable systems.

\subsection{Evolution of real space densities}\label{sec:realspace_densities}

\begin{figure*}[tbp]
  \includegraphics[width=0.90\linewidth]{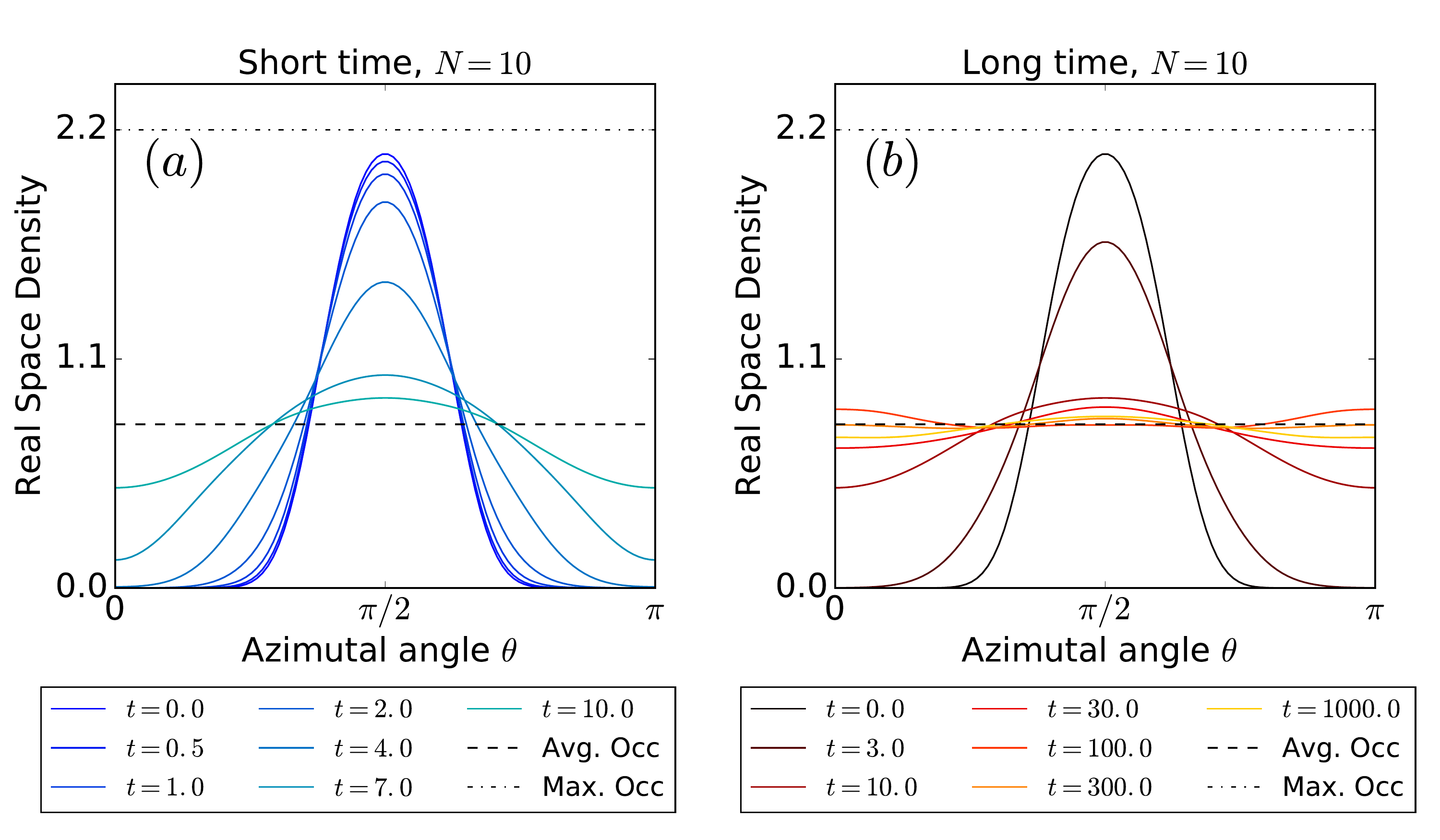}\\
\caption{Snapshots of the real space density distribution starting from the equatorial initial
  state, for $N=10$ electrons on a sphere.  Time slices are the same as in
  \figref{fig:orb_dens_evol}.  
At later times ($t\gtrsim30$) the density becomes approximately uniform, apart from persistent
fluctuations.  }
  \label{fig:real_space_dens_evol}
\end{figure*}

In \figref{fig:real_space_dens_evol} the dynamics are visualized through real-space density profiles.
The real-space profiles are obtained straightforwardly from the orbital occupancies $n_i$ using the
real-space wavefunctions $\psi_i(\vec{r})$ of the single-particle orbitals: the real-space density
at the location $\vec{r}$ is $\rho(\vec{r})=\sum_i n_i |\psi_i(\vec{r})|^2$.

For the system sizes we can reach, the initial profile (real-space profile of the initial state) is
a bell-shaped curve without a plateau (Sec. \ref{sec:initial_state}).  The density spreads out over
the full sphere and is eventually near-constant.

As noticed with the orbital density profiles, the density in-between the equator and poles relaxes
faster, while the densities at the equator and pole take longer to relax, and also perform
persistent long-term oscillations.

Remarkably, the non-monotonicity at short times in the orbital occupancy profile is washed out by
the orbital profiles and does not get reflected in the real space density profiles.  It is perhaps
reassuring that the widely used $V_1$ potential does not lead to artificial effects when viewed in
physical space, despite its singular (derivative of a delta function) form.

\subsection{Entanglement dynamics} \label{sec:entanglement}

\begin{figure*}[tbp]
  \includegraphics[width=.90\linewidth]{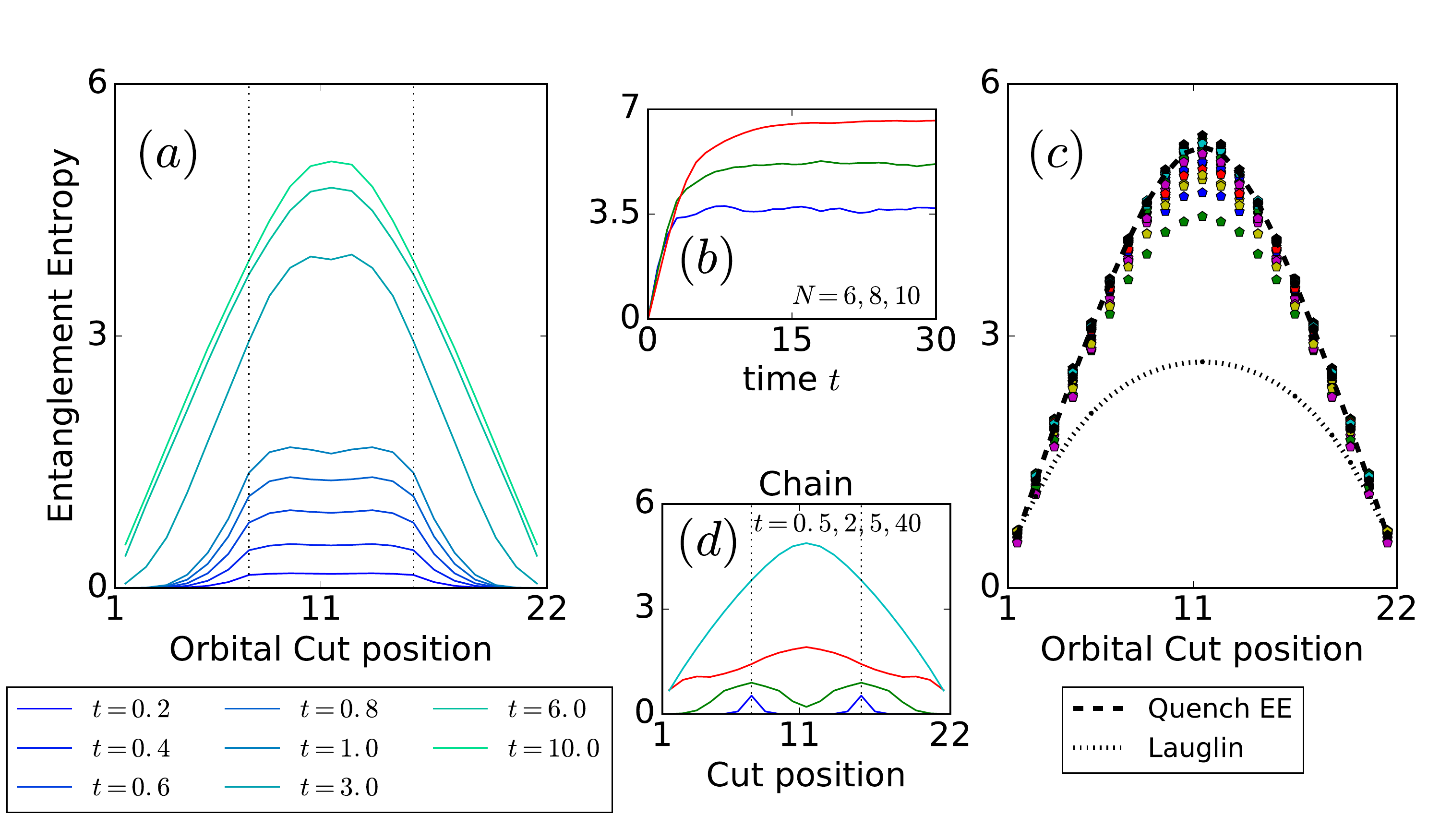}
  \caption{(a) Snapshots of orbital entanglement entropy profile, up to $t=10$.  Vertical lines show
    the boundaries of the initial equatorial state.  The entanglement entropy is saturated after
    $t\sim10$.
(b) Evolution of entanglement entropy between equal partitions (equatorial cut).
(c) Entanglement entropy profiles for the 20 eigenstates that are closest in energy to the
    expectation energy of the state.  The profile of the evolving state at a late time ($t=100$) is
    included for comparison, as well as the entanglement entropy of the Laughlin state.  
(d) Snapshots of entanglement entropy profiles for an open-boundary tight-binding chain with
    single-particle hopping and nearest-neighbor interaction, $H = -\sum(c_i^{\dagger}c_{i+1}
    +\mathrm{h.c.}) + \sum c_i^{\dagger}c_{i}c_{i+1}^{\dagger}c_{i+1}$, starting with an analogous
    initial state. 
  \label{fig:entanglement_entropy}
}
\end{figure*}

We now examine the entanglement generated in the dynamics.  

We calculate the bipartite entanglement entropy between two blocks of orbitals (conventionally
referred to as $A$ and $B$), the first partition containing the first block of $l_A$ orbitals
starting from one pole of the sphere and the second partition containing the rest of the orbitals.
This type of `orbital partitioning' entanglement has been widely studied since the beginning of
entanglement investigations of fractional quantum Hall systems, including studies of both
entanglement entropies \cite{HaqueSchoutens_PRL2007, Zozulya_PRB2007, FriedmanLevine_PRB2008,
  ZozulyaRegnault_PRB2009, MorrisFeder_PRA2009, LaeuchliBergholtzHaque_NJP10,
  ZaletelMongPollmann_PRL2013, LiuHaldaneSheng_PRB2016, LiuBhatt_PRL2016, LiuBhatt_PRB2017} and
entanglement spectra \cite{LiHaldane_PRL2008, ZozulyaRegnault_PRB2009,
  RegnaultBernevigHaldane_PRL2009, ThomaleRegnaultBernevig_PRL2010, LaeuchliBergholtzHaque_PRL2010,
  SterdyniakHaldane_NJP2011, ChandranHermannsBernevig_PRB2011, HermannsChandranBernevig_PRB2011,
  SterdyniakBernevig_PRL2011, LiuBergholtzLauchli_PRB2012, ZaletelMongPollmann_PRL2013,
  ZaletelMongPollmannRezayi_PRB2015, LiuKim_PRB2015, Regnault_arxiv1510, LiuHaldaneSheng_PRB2016}.
The orbital entanglement may be regarded as a reasonable proxy for entanglement between real-space
partitions of the sphere \cite{Bernevig_realspace_PRB2012, DubailRead_PRB2012,
  RodriguezSlingerland_PRL2012}, because of the sequential spatial arrangement of orbital positions.
(However, because the width of orbital wavefunctions is larger than their mutual spacing, this
correspondence is nontrivial.)
The interest in the orbital entanglement entropy/spectrum of ground state quantum Hall wavefunctions
is largely due to topological information contained in these quantities.  Topological aspects do not
play a role in the present work.  On the other hand, the evolution of entanglement is also widely
used to characterize non-equilibrium dynamics; this is the spirit of our analysis below.

\figref{fig:entanglement_entropy}(a) shows snapshots of the spatial profile of the entanglement
entropy, obtained by plotting the entanglement entropy against the cut position, $l_A$.  As we have
done until now, we focus on the equatorial initial state: the $N=8$ orbitals nearest to the equator
are filled, on a sphere with $3N-2=22$ orbitals.
At $t=0$ the entanglement entropy is zero, since the initial state is a product state.  The
entanglement entropy grows with time; the initial growth for almost all cuts in the equatorial
region is approximately linear.
This is seen explicitly in \figref{fig:entanglement_entropy}(b), where the entanglement entropy for
the half-partition ($l_A= \frac{1}{2}(3N-2)$) is plotted as a function of time.
After $t\sim10$ the entanglement entropy saturates.  The saturation time grows approximately
linearly with system size (\figref{fig:entanglement_entropy}(b)) and correlates with the time it
takes for the quantum fluid to reach the north and south pole of the sphere
(\figref{fig:orb_dens_evol}).

At long times, the entanglement entropy profile eventually settles to values consistent with the
entanglement entropy in high-energy eigenstates, or equivalently, with the entanglement entropy in
random or `thermal' eigenstates \cite{Deutsch_NJP2010, SantosPolkovnikovRigol_PRE2012,
  SharmaDeutsch_PRE2013, Alba_PRB2015, BeugelingHaque_JSM2015, GarrisonGrover_arxiv2015,
  Watanabe_arXiv2017, VidmarRigol_PRL2017}.  This is seen in the right panel, where we have plotted
the orbital entanglement entropy profiles (entanglement entropy versus $l_A$) for the 20 eigenstates
that are closest in energy to the expectation energy of the time-evolving (or initial) state.  The
long-time entanglement profile follows the eigenstate entanglement profile quite closely.

It is interesting to compare the long-time (or eigenstate) entanglement profile with the ground
state entanglement entropy of the Laughlin state, studied first in detail in
Ref.\ \cite{HaqueSchoutens_PRL2007}.  The ground state entanglement entropy follows an area law
\cite{Srednicki_arealaw_PRL1993, Hastings_PRB2007, Masanes_PRA2009, Eisert_arealaw_RMP2010} and
encodes topological properties through its subleading term, the so-called topological entanglement
entropy \cite{LevinWen_PRL2006, KitaevPreskill_PRL2006}.  The long-time entanglement entropy after a
quench, in contrast, should follow a volume law, and is more similar to a typical random state than
to a low-energy state.  Indeed the long-time entanglement profile is seen in
\figref{fig:entanglement_entropy}(c) to be significantly larger than the Laughlin (ground state)
profile.  Distinguishing volume law from area law quantitatively in finite size data however
requires careful finite-size scaling.  For the quantum Hall geometry on the sphere, area law and
volume law behavior mean that the top of the entanglement profile scales as $\sim \sqrt{N}$ and as
$\sim N$ respectively.  Our quench data, \figref{fig:entanglement_entropy}(b), shows that the
entanglement entropy for the half-system cut indeed saturates at values growing linearly with system
size.

We now return to the short-time behavior, which displays a striking effect of the nature of our
Hamiltonian.  At very short times, the entanglement entropy profile is flat in the equatorial regime
where the product state fermions are initially placed --- there is no peak at the two edges.  This
is in sharp contrast to what one finds in the case of regular (uncorrelated) hopping, as we display
in \figref{fig:entanglement_entropy}(d), where entanglement profiles are shown for a tight-binding
chain starting from the corresponding initial state --- 8 fermions placed in the middle of a 22-site
chain.  (The chain Hamiltonian is that given in the Introduction, Eq.\ \eqref{eq:H_tightbinding},
with open boundary conditions and $V=1$.)  An uncorrelated-hopping Hamiltonian, with hopping
terms of the form $c^{\dagger}c$ rather than $c^{\dagger}c^{\dagger}cc$, generates entanglement
first at the boundaries between filled and unfilled regions.  The flat entanglement profile
generated in our case is a consequence of the correlated hopping processes
governing the dynamics in the lowest Landau level Hamiltonian.

With correlated hopping, a fermion hops from the initial central region to an initially empty site
only when there is a partner fermion hopping across the other boundary of the initial filled region.
According to intuitive ideas of entanglement, this correlated process clearly will generate
entanglement between the two edge regions, and will show up as nonzero entanglement across a cut at
the center.  Thus correlated hopping automatically generates entanglement across a cut placed
between the two edges; hence the flat profile at very short times.  In contrast, for uncorrelated
hopping, a hop on the left edge and a hop at the right edge are independent; these processes at
short times generate entanglement across cuts near either edge but no entanglement across cuts
halfway between the edges.  Thus the short-time entanglement for a tight-binding chain has separated
peaks near either edge, as seen in \figref{fig:entanglement_entropy}(d).

This intuitive picture can be made more precise through a toy calculation.  Using the notation
$\ket{0^n1^{l}0^n}$ for our initial state (a block of $l=N$ fermions sandwiched between two
blocks of $n$ empty sites), a tight-binding nearest-neighbor uncorrelated hopping Hamiltonian
generates the following wavefunction at linear order in time:
\begin{equation*}
\mathcal{N}_1 \Big( \ket{0^n1^{l}0^n}+ \alpha\ket{0^{n-1}101^{l-1}0^n}  +
  \alpha\ket{0^{n}1^{l-1}010^{n-1}}     \Big) 
\end{equation*}
where $\alpha$ is a small parameter proportional to time, and
$\mathcal{N}_1=\left(1+2|\alpha|^2\right)^{-1/2}$
normalizes the wavefunction. With this simplified wavefunction, for a cut between the edges we obtain
the reduced density matrix for one partition to be
$\rho_A=\mathcal{N}_1^2
\left(\begin{array}{cc}  1+|\alpha|^2 & \alpha \\  \alpha^* &  |\alpha|^2 \end{array}\right)$.
This leads to \emph{zero} entanglement at leading order (order $|\alpha|^2$).  A cut at either edge,
on the other hand, leads to finite entanglement at leading order.  This explains the double-peak
structure at small times in \figref{fig:entanglement_entropy}(d).
  
To model the correlated hopping effect tractably, we make the simplification that correlated nearest-neighbor hoppings ($m=1$ terms in \figref{fig:hopping}) dominate.
The short-term wavefunction is then 
\begin{equation*}
\mathcal{N}_2 \Big( \ket{0^n1^{l}0^n}+ \gamma\ket{0^{n-1}101^{l-2}010^{n-1}} \Big) 
\end{equation*}
where $\gamma$ is a small parameter proportional to time, and $\mathcal{N}_2 =\left(1+|\gamma|^2\right)^{-1/2}$.
Here, hoppings across the two edges do not generate separate terms, but are correlated.  
Now, for a cut between the edges we obtain the reduced density matrix
$\mathcal{N}_2^2 \left(\begin{array}{cc}1 & 0\\  0 & |\gamma|^2 \end{array}\right)$,
yielding a nonzero entanglement entropy at leading order, $\mathcal{O}\left(|\gamma|^2\right)$.
A cut at the edge yields an identical entanglement entropy at leading order.
This explains the flat nonzero entanglement profile in the region between the edges for the lowest-Landau-level Hamiltonian.

\subsection{Lopsided initial state} \label{sec:Lopsided}

\begin{figure*}[tbp]
\includegraphics[width=0.90\linewidth]{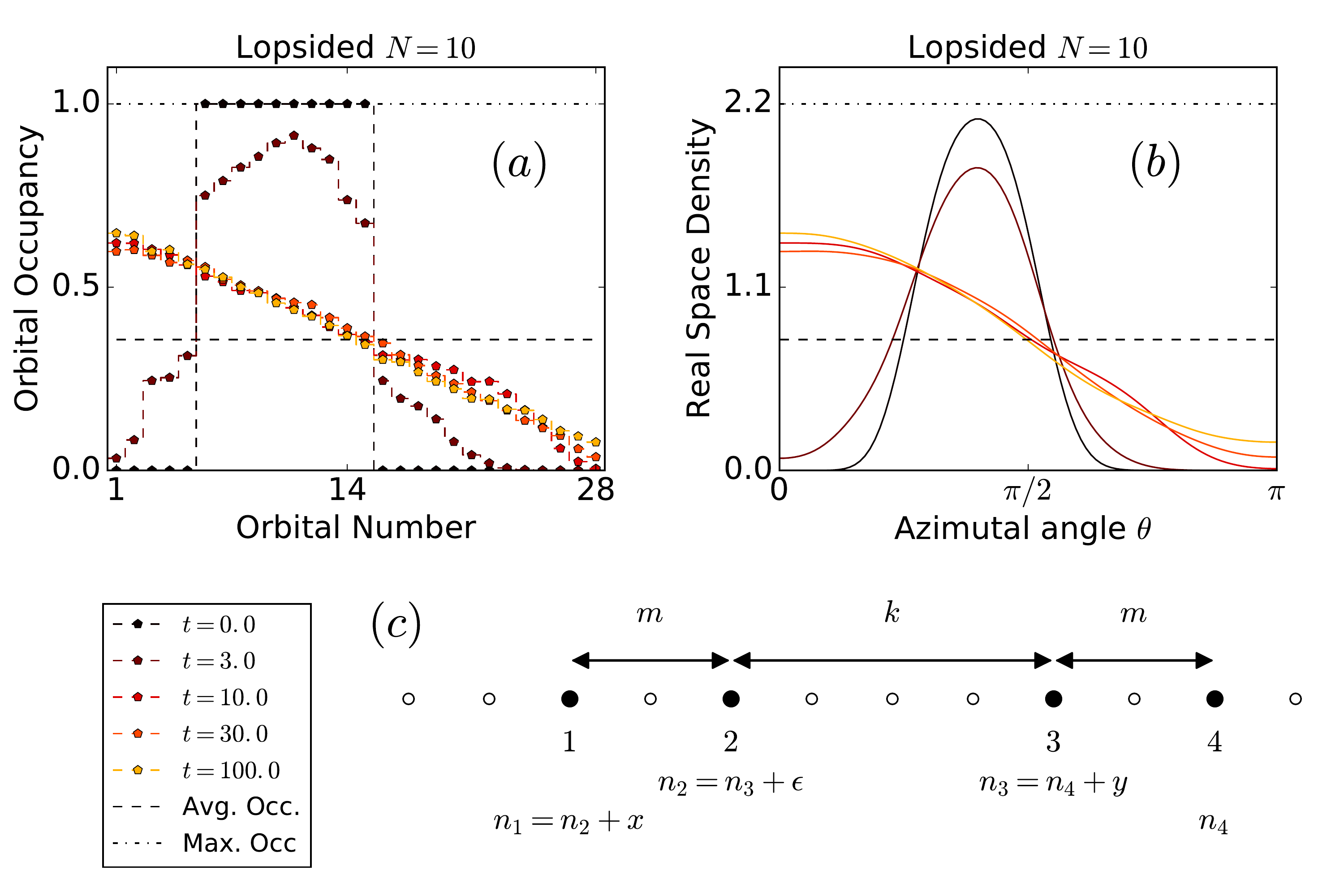}
\caption{Snapshots of the (a) orbital occupancy profile and (b) real space density distribution.
  Starting from a lopsided state; $N=11$ electrons on a sphere.  At $t\sim10$ the profile has
  approximately reached its asymptotic form.  (c) Notation used for mean-field-like model in text.
}
\label{fig:lopside}
\end{figure*}

Until now, we have described dynamics starting from the equatorial initial state, $\ket{0^l1^N0^l}$,
with $l\approx N$, where the filled block of orbitals is placed symmetrically around the equator.
In this case the total angular momentum is zero, $L_z=0$.  Since $L_z$ corresponds roughly to total
`latitude' (angular position) on the sphere and is conserved, the state at any time is also
symmetric around the equator.  The long-time state, having uniform filling on all orbitals, is
obviously symmetric as well.

We now consider initial states where the block is displaced from the symmetric location.  Taken to
the extreme, this results in the polar block (maximal $L_z$) which we have explained is an inert
state (Section \ref{sec:initial_state}).  We now consider initial states intermediate between these
two cases, such that $L_z$ is nonzero but not maximal.  Since the time evolution conserves $L_z$,
the state continues to be lopsided and biased toward one pole.

We first examine moderate displacements from the symmetric situation, $\ket{0^{l_1}1^N0^{l_2}}$,
with $l_1\approx \frac{1}{2}N$ and $l_2\approx \frac{3}{2}N$.  Results for such a case are shown in
\figref{fig:lopside}(a,b).  As explained above, the density profile remains lopsided due to $L_z$
conservation --- the long-time asymptotic profile is not uniform.  This is true for both the orbital
occupancy and the real-space density.

The asymptotic occupancy profile in \figref{fig:lopside}(a) is, remarkably, nearly linear.  While
$L_z$ conservation guarantees an inhomogeneous profile, it does not by itself predict this simple
shape for the final profile.  We can interpret the linear shape using a `hydrodynamic' or `rate
equation' approach for the correlated hopping term.  Let us consider four orbitals such that
correlated hopping can occur from the outer two orbitals to the inner two, or vice versa.  The
situation is depicted in \figref{fig:lopside}(c) as the four filled dots.  If the orbitals are
labeled 1, 2, 3, and 4, then orbitals 1 and 2, as well as orbitals 3 and 4, are separated by $|m|$,
so that the hopping term $V_{km}$ can transfer a pair of fermions from the outer orbitals 1 and 4 to
the inner orbitals 2 and 3, or (by hermiticity) from the inner orbitals 2 and 3 to the outer
orbitals 1 and 4.  The long-time steady state (`equilibrium') densities should be such that the
$(14)\to(23)$ rate matches the $(23)\to(14)$ rate.  A mean-field or hydrodynamics-like description
of this condition is
\begin{equation}  \label{hydro1}
n_1n_4 (1-n_2)(1-n_3) ~=~ n_2n_3 (1-n_1)(1-n_2) .
\end{equation}
If we parametrize the differences of densities as
\begin{equation}
n_1 = n_2+x  = (n_3+\epsilon)+x  = (n_4+y) + \epsilon + x , 
\end{equation}
then a linear occupancy profile would correspond to $x=y$, because the 1--2 and 3--4 distances are
equal.  We can work in the limit where $x,y,\epsilon$ are much smaller than $n_i$, \ie the four
orbitals are not very far apart.  At first order in this approximation, the rate equation
\eqref{hydro1} yields
\begin{equation}
n_4^2(1-3\epsilon-2x-4y) +n_4(x+\epsilon+y) = n_4^2(1-3\epsilon-x-5y) +n_4(\epsilon+2y)
\end{equation}
so that we obtain $x=y$.  Thus the approximate hydrodynamic description predicts a linear occupancy
profile.  (The difference $\epsilon$ is not determined by this calculation, which is consistent as
the argument is independent of the distance $k$.

Near-linearity of the final distribution is a nontrivial consequence of the symmetric correlated
hopping, coupled with the geometry allowing for a hydrodynamic description.  Applying the same type
of argument to a non-correlated single-particle hopping term (between orbitals 1 and 2) gives
$n_1(1-n_2)=n_2(1-n_1)$, \ie $n_1=n_2$, a flat density profile.  Of course, it is difficult to
imagine an analog of $L_z$ conservation with a single-particle hopping term; thus a non-homogeneous
asymptotic state is not expected.

\begin{figure*}[tbp]
\includegraphics[width=0.90\linewidth]{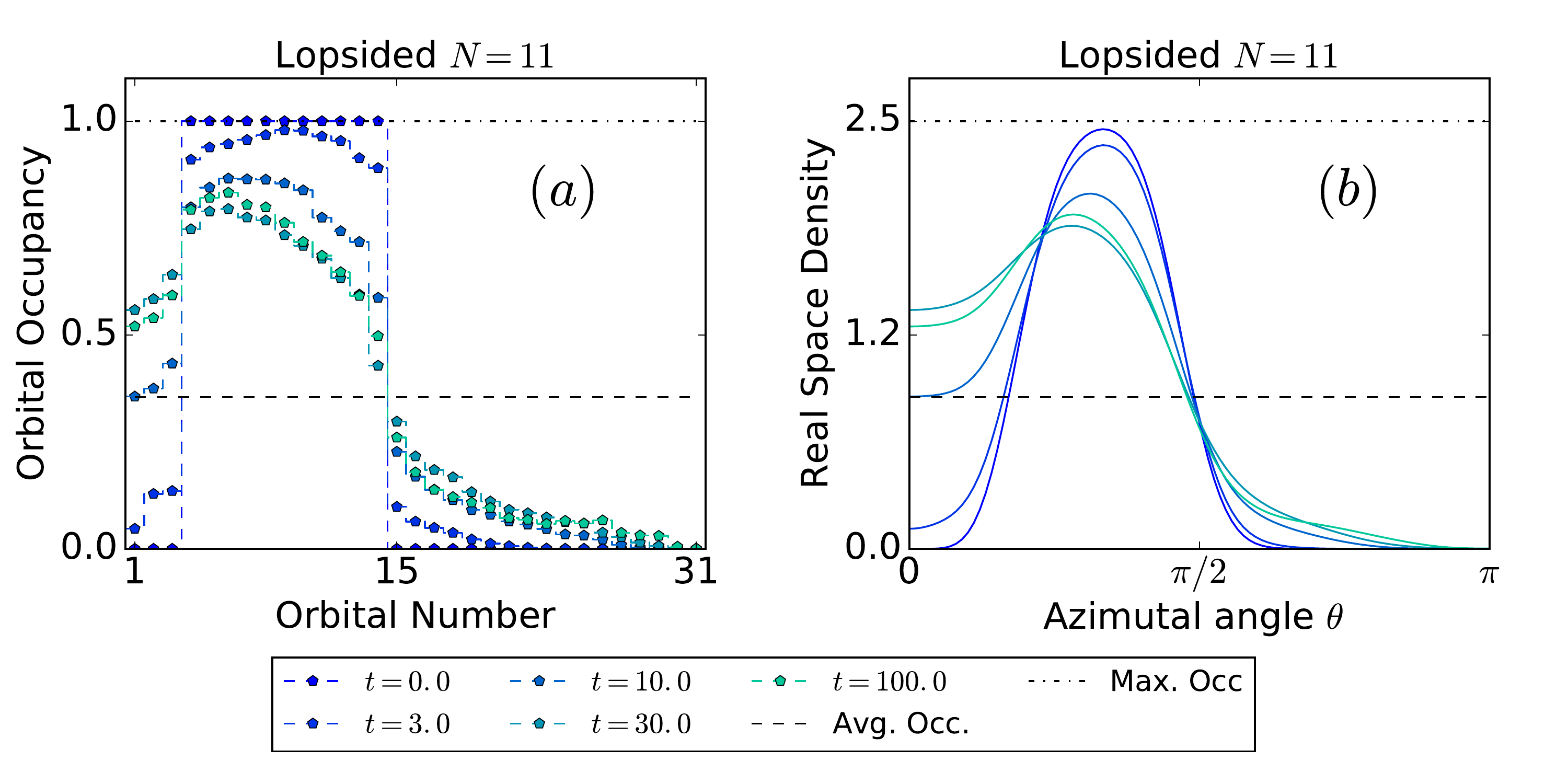}
\caption{Snapshots of the (a) orbital occupancy profile and (b) real space density distribution, for
  $N=11$ electrons on a sphere. 
The  initial state is even more lopsided than in \figref{fig:lopside}. 
  }
\label{fig:morelopside}
\end{figure*}

By squeezing the initial block of fermions closer to the pole, we can remove the geometric freedom
required for a mean-field description.  This is shown in \figref{fig:morelopside}.  This may be
regarded as an intermediate case between the moderately lopsided case where the occupancy profile
approaches a linear form, and the polar initial state for which the density profile never changes.
In the present case, there is some dynamics (the Hilbert space is smaller than equatorial or
moderately lopsided cases, but nevertheless finite).  However, the long-time state now has a strong
remnant of the initial peak in the occupancy profile.

\subsection{Overlap distribution and torus dynamics \label{sec_overlapdist}}

\begin{figure*}[tbp]
  \includegraphics[width=.90\linewidth]{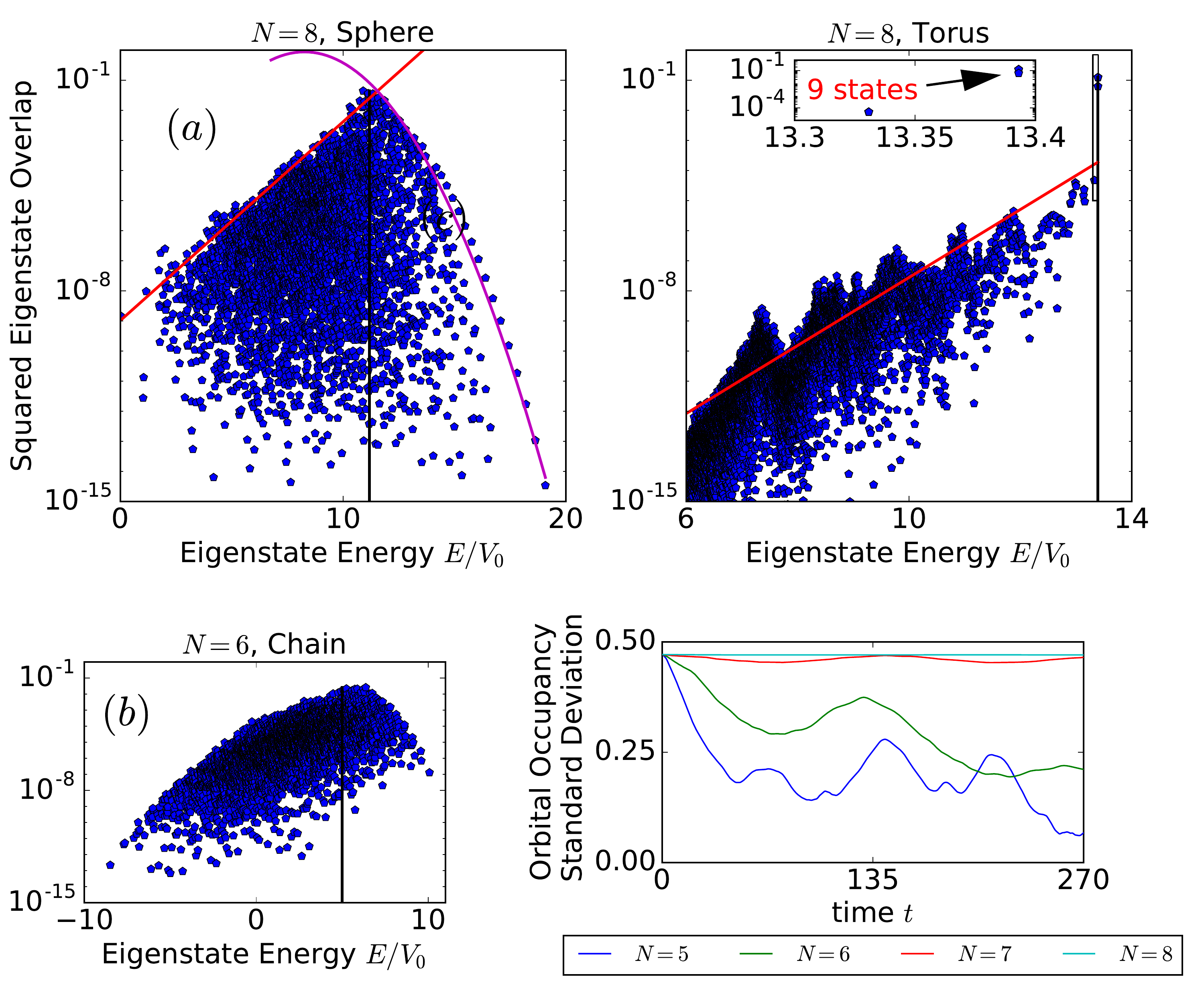}
  \caption{
(a) Overlaps of equatorial initial state with energy eigenstates, $N=8$ fermions on a
    sphere. 
(b) Overlaps of analogous initial state $\ket{0^l1^N0^l}$ with  energy eigenstates, $N=6$
    fermions on a tight-binding chain.  (Defined in caption to \figref{fig:entanglement_entropy}.)
(c) Overlaps of block initial state \eqref{eq:initial_block_state} for $N=8$ fermions on a torus. 
Inset zooms into highest-energy eigenstates. 
In (a,b,c), black vertical line shows energy of initial state.  Red/purple lines in (a,c) are guides to the eye.
(d) Dynamics on torus, displayed through the standard deviation of orbital occupancies, as in
\figref{fig:variance}.  The dynamics is nearly frozen, increasingly so for larger systems.
 \label{fig:overlaps}
}
\end{figure*}

The dynamics of a quantum state is governed by the eigenstates of Hamiltonian with which the state
has significant overlap.  The distribution of overlaps --- the overlaps with eigenstates as a
function of corresponding eigenenergies, thus has been examined in the non-equilibrium literature
for various Hamiltonians and initial states \cite{RigolOlshanii_Nature2008, Rigol_PRA2009,
  BiroliKollathLauchli_PRL2010, CassidyRigol_PRL2011, RigolFitzpatrick_PRA2011,
  SantosPolkovnikovRigol_PRL2011, SorgPolletHeidrichM_PRA2014, Shchadilova_PRL2014,
  MazzaHaque_JSM16, Motrunich_PRA2017, AbaninPapic_arxiv2017}.

In \figref{fig:overlaps}(a) we show the overlaps of the equatorial initial state with the
eigenstates of the Hamiltonian.  We consider overlaps with all eigenstates in the $L_z=0$ sector.
Eigenstates with any $L^2$ can have nonzero overlap with the equatorial initial state.  The energy
of the equatorial state, indicated with a black vertical line, is far from the edges of the
spectrum, but nearer to the upper end of the spectrum than the lower end.  The highest overlaps come
from eigenstates in this region, but not all eigenstates in this region have a large overlap.  The
largest overlaps in other energy regions follow an approximately exponential envelope, $\sim
e^{\lambda (E-E_{\rm init})}$, for $E<E_{\rm init}$, and an approximately Gaussian envelope, $\sim
e^{-(E-E_{\rm init}+\lambda)^2/\sigma^2}$, for $E>E_{\rm init}$.

The overlap distribution is similar (in terms of overall shape and exponential decrease) to the
overlap distribution for the tight-binding interacting chain, shown for comparison in
\figref{fig:overlaps}(b).  (This is the same chain of interacting fermions with single-particle
uncorrelated hopping, used previously in \figref{fig:entanglement_entropy}.  The initial state is
$N$ fermions in the central $N$ sites among $3N-2$ sites.)  This similarity suggests that the
overall shape and the exponential envelope is not due to the correlated-hopping part of the
Hamiltonian.  Rather, it is due to the static interaction part ($m=0$ terms of $V_{km}$), which
prefers to keep the fermions apart from each other.  Lower energy eigenstates have smaller
contributions from configurations with large `static' interaction energy.

The Gaussian envelope at larger energies follows a well-distinguished set of eigenstates.  Such
eigenstate branches appear to be common in overlap distributions, \eg in Figure 5a of
\cite{MazzaHaque_JSM16} and in Figure 3a of Ref.\ \cite{AbaninPapic_arxiv2017}.  In the present
case, states in this branch have much smaller overlaps than the dominant eigenstates, and hence do
not play any noticeable role in the dynamics, unlike the case of Ref.\ \cite{AbaninPapic_arxiv2017}.

In \figref{fig:overlaps}(b) we show the overlap distribution for the analogous initial state on the
torus.  The dominant eigenstates are now the few (around $N$) highest-energy eigenstates.  These
eigenstates have the structure of linear combinations of product states like our initial state,
centered at different positions of the torus.  These few eigenstates become increasingly dominant
with increasing system size.  A result of this extreme overlap distribution is that the dynamics is
frozen.  This is shown in \figref{fig:overlaps}(d).  The effect is more extreme for larger sizes.

Since the static part of the interaction energy is minimized by maximizing the spacing between the
fermions, it is expected that our spatially packed initial states should have energy nearer to the
top of the spectrum.  On the torus, this type of state appear at the very top of the spectrum.  On
the sphere, one can create states with energies even higher than our equatorial initial states.
(For example, since packed configurations near the poles are more costly than packed configurations
near the equator, we can pack $N/2$ fermions near the north pole and the rest near the south pole,
obtaining an $L_z=0$ configuration with energy higher than the equatorial state.)  This explains why
the energy of the equatorial initial state in \figref{fig:overlaps}(a) is not at the very top of the
energy spectrum.

\section{Level statistics --- Integrability and universality classes}\label{sec:level_statistics}

We now consider the level spacing statistics of the fermionic lowest-Landau-level Hamiltonians.

Considerations of level statistics pervade the quantum dynamics literature, both in the field of
single-particle quantum chaos \cite{BerryTabor_ProcRSoc1977, BohigasEA_PRL1984, Haake_2010_Book,
  Wimberger_2014_Book} and in the study of many-body quantum dynamics
\cite{PolkovnikovRigol_AdvPhys2016}.  The main reason is that integrability manifests itself in the
level spacing distribution, and quantum dynamics can be strongly affected by the presence or
proximity of integrability.  The spectrum of an integrable system has Poissonian spacing statistics,
whereas the spectrum of a non-integrable system will show Wigner-Dyson (GOE, GUE or GSE) statistics
\cite{BerryTabor_ProcRSoc1977, BohigasEA_PRL1984, Haake_2010_Book, Wimberger_2014_Book}.  Of course,
it is assumed that the energy spectrum is first sorted according to symmetry sectors.  One can thus
check for integrability using an analysis of level statistics.  As far as we are aware, no study of
the level statistics of FQH systems has been published to date. We present such a study now.  Our
analysis shows explicitly that the LLL-projected Hamiltonians are not integrable in the geometries
considered.  Since a magnetic field violates time reversal symmetry, GUE (Gaussian Unitary Ensemble)
statistics would be the obvious expectation, as GUE is generally associated with the lack of time
reversal invariance.  However, we show and explain below that GOE statistics appears for the sphere
and for several sectors of the torus.

\begin{figure}[tbp]
  \includegraphics[width=0.9\linewidth]{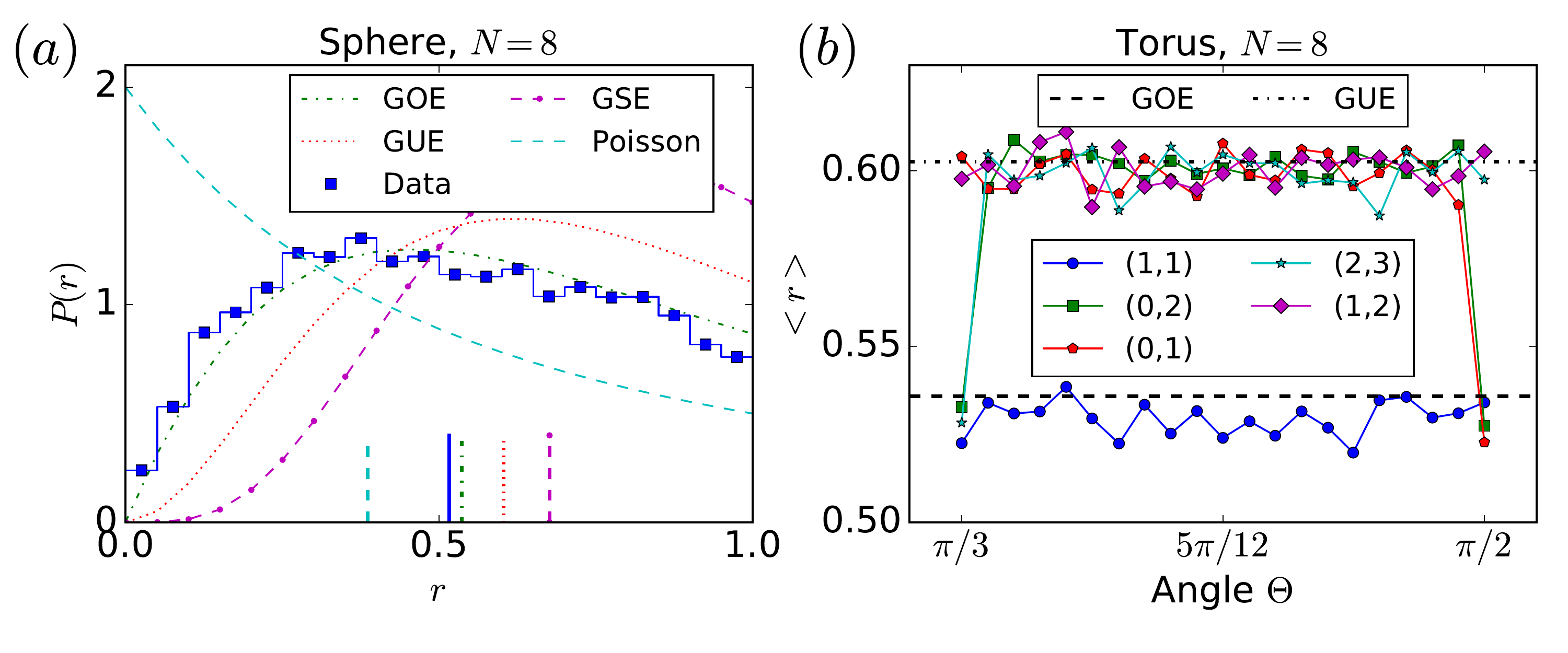}
\caption{
(a) 
$r$-statistics for the spectrum of the Hamiltonian on the sphere with $N=8$ fermions.  The dashed
  lines are expected $r$-distributions for different (random) matrix classes
  \cite{AtasBogomolnyRoux_rstat_PRL2013}.  The solid vertical lines show averages,
  $\langle{r}\rangle$, for the different classes and for the data.  The plot shows that the sphere
  Hamiltonian is in the GOE class.  The $r$-values are collected by computing the level spacings in
  each $L^2$ sector individually and then aggregating the different $L^2$ sectors.  No trimming has
  been applied.
(b) 
Averages, $\langle{r}\rangle$, for toroidal geometries ($\tau=e^{i\theta}$) with $N=8$ fermions in
selected momentum sectors.  $\theta$ is tuned from $\theta=\pi/3$ (hexagonal) to $\theta=\pi/2$
(square).  Most momentum sectors are in the GUE symmetry class for generic $\theta$, the only
exception being momentum sectors $(K_1,K_2)$ where $K_1=\pm K_2$, see also \figref{fig:Torus_rstat}.
Away from these special sectors $(K,\pm K)$, GOE statistics is only obtained for high-symmetry
geometries such as the square or hexagon.
}
\label{fig:r_stat_scan}
\end{figure}

\begin{figure*}[tbp]
  \begin{center}
    \includegraphics[width=0.90\linewidth]{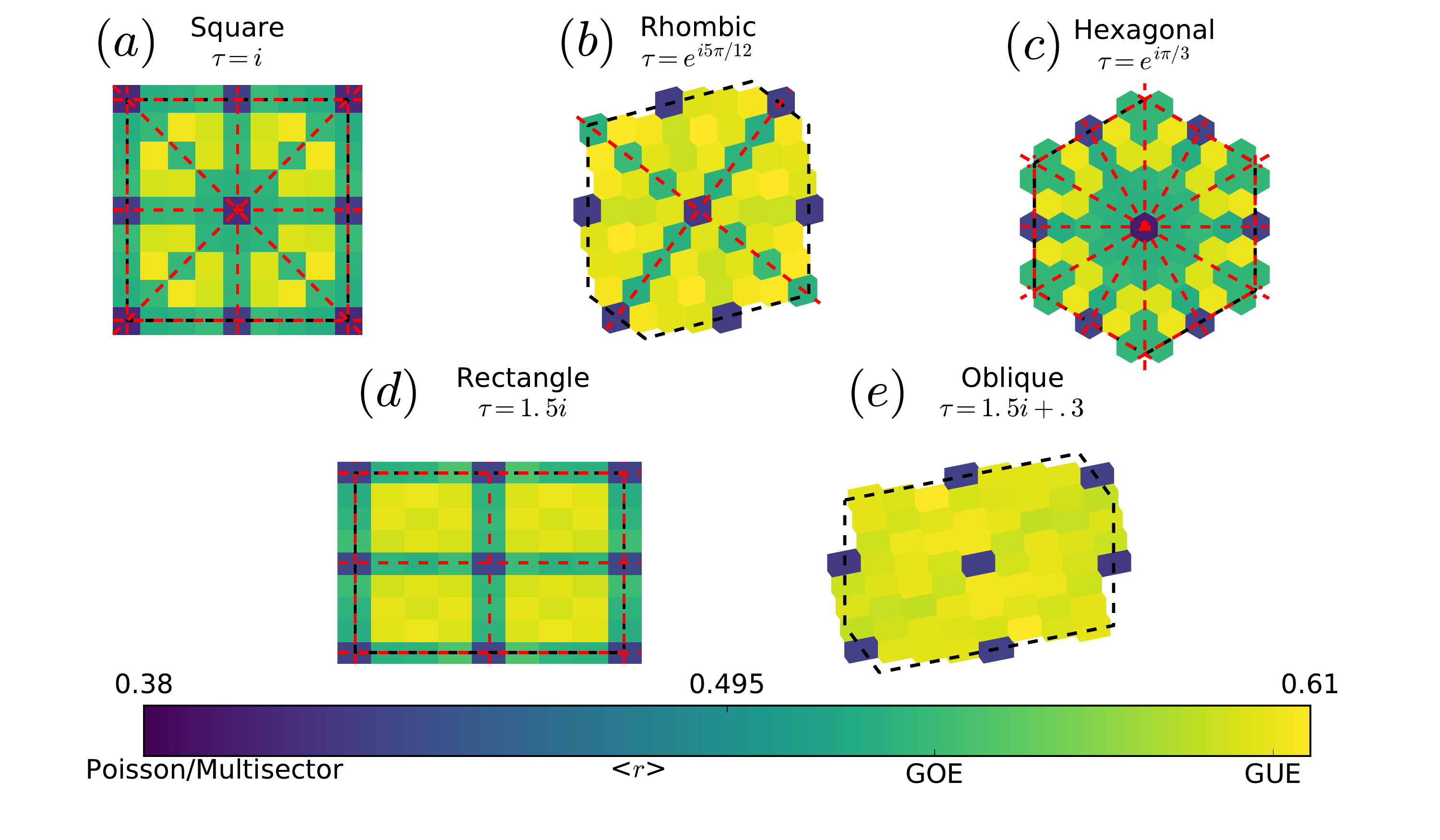}
  \end{center}
  \caption{
    $\langle{r}\rangle$ values for the $N=8$ spectrum in a selection of toroidal geometries.
    Each colored square/hexagon is a momentum sector $(K_1,K_2)$.
    Dashed black lines show the borders of the many body Brillouin zone.  
    The darkest blue sectors are $(K_1,K_2)=(0,0),(N/2,0),(0,N/2),(N/2,N/2)$, the latter three existing only when $N$ is even.
    Here there is a residual symmetry of orbital inversion, which superficially makes the statistics look Poissionian ($\langle{r}\rangle\approx0.386$, purple).
    In the other sectors, the $\langle{r}\rangle$ value is clearly identifiable as either GOE ($\langle{r}\rangle\approx0.536$, green) or GUE ($\langle{r}\rangle\approx0.603$, yellow).
    GOE statistics is found in momentum sectors which are invariant under the product of mirror and
    time-reversal symmetry.
    These sectors are located along the mirror axes of each geometry.
    All other momentum sectors have GUE statistics.
    To illustrate this effect, we have represented tori corresponding to all possible Bravais lattices for the system size $N=8$.
    The red dashed lines are reflection axes.  
  \label{fig:Torus_rstat}
}
\end{figure*}

For the LLL-projected Coulomb Hamiltonians, there is no \emph{a priori} reason to expect
integrability.  However, it is not easy to rule out the possibility, as the projection to the LLL is
a nontrivial operation.  (It is well known that projection onto Wannier levels can drastically
affect integrability properties, \eg the continuum Lieb-Liniger model is integrable while the
lattice Bose-Hubbard model is not.)  There is also the possibility that the pseudopotential ($V_1$)
Hamiltonian could be integrable in one of the geometries.  Also, it is not clear that
(non-)integrability in one of the common geometries (sphere, torus) implies the same for the other.
It is worth noting that the similar low-energy properties of these different situations (Coulomb and
$V_1$, sphere and torus) do not imply that integrability properties are the same.  To our knowledge,
the integrability of LLL Hamiltonians has not until now been examined explicitly.
In our analysis described below and displayed in \figref{fig:r_stat_scan} and
\figref{fig:Torus_rstat}, we show that each symmetry sector of the sphere and the torus displays
Wigner-Dyson statistics, for the filling appropriate for the Laughlin-1/3 state.  We display these
results for the $V_1$ potential that we have focused on in this work, but we have also checked that
the same holds true for the Coulomb potential.  (In fact, there is no difference between the $V_1$
and Coulomb potentials with respect to the spectral statistics, for all examples we have checked.)
This confirms that the LLL Hamiltonians under question are non-integrable.

We characterize the level statistics through calculations of the 
ratio of consecutive level spacings $\langle r\rangle$
\cite{OganesyanHuse_rstat_PRL2007, AtasBogomolnyRoux_rstat_PRL2013}. This
measure is based on $s_n = E_{n+1}-E_n$, the set of level spacings in
an ordered list ${E_n}$ of eigenenergies. The ratio
\begin{equation}
r_n = \frac{{\rm min}(s_n, s_{n-1})}{{\rm max}(s_n, s_{n-1})}
\end{equation}
is defined for each pair of consecutive level spacings.  The
statistics of the ratio $r_n$ is more convenient than the bare level
spacings $s_n$ because it bypasses the need to account for varying
density of states through unfolding procedures.  

For Poisson statistics, the probability distribution of $r_n$ is $P(r)=2/(1+r)^2$ with mean $\langle
r \rangle=2\ln 2-1 \approx 0.39$.  For the Wigner-Dyson ensembles, the probability distributions
are well-approximated by the surmise \cite{AtasBogomolnyRoux_rstat_PRL2013} $P(r) \propto
(r+r^2)^{\beta}/(1+r+r^2)^{1+3\beta/2}$ up to normalization, with $\beta=1$ ($\beta=2$) for GOE
(GUE).  The averages are $\langle r \rangle_{\rm GOE}\approx 0.53$ and $\langle r \rangle_{\rm
  GUE}\approx 0.60$ \cite{AtasBogomolnyRoux_rstat_PRL2013}; these values provide a quick check for
the nature of numerically obtained spectra.

\subsection{Sphere}

In \figref{fig:r_stat_scan}(a) we show the distribution of $r$ values for the spectrum of the $V_1$
Hamiltonian on the sphere.  One has to first separate the symmetry sectors.  Numerical
diagonalization is naturally done in distinct $L_z$ sectors.  We consider here the $L_z=0$ sector
which contains the equatorial initial state and the Laughlin state.  However a single $L_z$ sector
contains multiple $L^2$ sectors.  The $L_z=0$ spectrum was post-processed into separate spectra for
each $L^2$ sector (by numerically applying the $L^2$ operator on the eigenstates), and the $r$
values for each sector were collected separately.  Under the assumption that each $L^2$ sector
possesses the same statistics, it is reasonable to combine the groups of $r$ values and present the
combined distribution, as we have done.  Combining the level spacings ($s$ values) themselves from
different sectors would have been far trickier as one has to take into account the different density
of states effects in the different sectors.  Considering the ratios $r$ thus allows us to obtain
reasonable statistics even when the individual sector Hilbert spaces are quite modest-sized.  The
distribution $P(r)$ obtained in this manner very clearly follows the GOE form.  The excellent
agreement provides \emph{a posteriori} justification for combining data from different $L^2$ sectors
--- if different sectors had different statistics, one would expect the numerical $P(r)$ histogram
to be intermediate between two or more of the standard distributions.

We have thus shown that the sphere Hamiltonian has GOE statistics. (We have checked that GOE
statistics is also obtained in other $L_z$ sectors.)  Level statistics following one of the
Wigner-Dyson distributions demonstrates the Hamiltonian is non-integrable.  To explain why the
statistics is GOE rather than GUE, one has to consider symmetries of the system.  The conventional
expectation is that Hamiltonians with time-reversal symmetry have GOE statistics, while those with
broken time-reversal symmetry have GUE statistics.  In most usual cases, the Hamiltonian matrix
elements are real in the first situation and complex in the second situation.  This latter
distinction however, cannot obviously be rigid as any complex Hamiltonian matrix can be
basis-transformed to have real matrix elements, and vice versa.  Clearly, since we have GOE
statistics in a situation with a magnetic field (which breaks time-reversal symmetry), the current
situation is more involved than the simple picture above.

The explanation is that, if the Hamiltonian breaks time-reversal symmetry but is invariant under
another anti-unitary transformation (\eg time-reversal coupled with a reflection), then the
statistics is GOE.  This was first pointed out by Robnik and Berry \cite{RobnikBerry_JPA1986} in the
single-particle context.  To our knowledge, this is the first time a many-body example has been
discussed.  In the case of the Haldane sphere, a combination of time-reversal and a reflection
around the equatorial plane keeps the system invariant.  Time-reversal reverses the directions of
cyclotron motion along each orbital, \ie it changes the sign of $L_z$ on each orbital without
changing the position, so that the orbitals in the northern hemisphere now have negative $L_z$.
(This is equivalent to changing the sign of the monopole at the center of the sphere.)  A reflection
around the equatorial plane switches the positions of the orbitals.  Thus the combination of the two
operations is an anti-unitary operation that keeps the system invariant.  In this manner, we obtain
GOE statistics in a many-body system violating time-reversal mechanism, through the mechanism of
Ref.\ \cite{RobnikBerry_JPA1986}.

\subsection{Torus}

In \figref{fig:r_stat_scan}(b) we summarize some results for the various torus geometries.
A different distribution appears in each momentum sector (as defined in \ref{sec:symmetry_sectors}).
Now, we only show the average $r$ values and omit the full distributions.
Several representative momentum sectors are shown as a function of the angle $\theta$
(following the parametrization introduced in \ref{ref_singlepcle_torus}).
The numerical diagonalizations are performed in individual momentum sectors, so no post-processing is necessary.
(Both momentum symmetries are exploited in the numerical diagonalization.)
The $\langle{r}\rangle$ values are close to either the GOE or the GUE values for every momentum sector shown in \figref{fig:r_stat_scan}(b), and for every torus orientation.
This demonstrates that the Hamiltonians are non-integrable.
Whether $\langle{r}\rangle$ is close to GOE or GUE expectation reflects the presence or absence of an anti-unitary combination of operators which preserves the Hamiltonian in that sector.

The torus sectors are presented in more detail in \figref{fig:Torus_rstat}.  In the sectors
$(K_1,K_2)= (0,0)$, $(N/2,0)$, $(0,N/2)$, $(N/2,N/2)$, there is an additional symmetry (orbital
inversion) not resolved in the numerical diagonalization.  The combination of spectra from two
symmetry sectors results in a combined spectrum which has no level repulsion, and the resultant
distribution looks closer to Poissonian than Wigner-Dyson.  This is reflected in the corresponding
squares/hexagons in \figref{fig:Torus_rstat} having near-Poissonian values of $\langle{r}\rangle$.

The Hamiltonian in most $(K_1,K_2)$ sectors has GUE statistics, reflecting the broken time-reversal
symmetry due to the magnetic field.
However, a smaller number of momentum sectors has GOE statistics in spite of the broken time-reversal symmetry:
for example the momentum sectors $K_1=\pm K_2$ for tori of aspect ratio $1$.
In this case, time-reversal combined with a reflection along the diagonal keeps the Hamiltonian invariant.
More generically, we note that the momentum sectors with GOE statistics are determined by the Bravais lattice symmetry of the considered geometry.
These sectors indeed coincide with the reflection axes of each torus,
such that the product of time-reversal and reflection leaves the Hamiltonian unchanged in these sectors.
To verify this statement, we have obtained the level-spacing statistics in geometries representative of all possible types Bravais lattices.
The results for the largest size ($N=8$) are given in Fig.~\ref{fig:Torus_rstat}.

A noteworthy feature is that GOE and GUE statistics do not correspond to real and complex Hamiltonians. 
For the sake of concreteness, let us consider an example in the square geometry.
In that case, the sectors $(K_1, K_2) = (K, \pm K)$ ($K \neq 0$) have GOE statistics even though the Hamiltonian has complex matrix elements.

\section{Discussion and Context} \label{sec:concl}

We have presented a detailed exploration of non-equilibrium dynamics and level-spacing statistics
of an interacting fermion system in the lowest Landau level,
on the commonly used sphere and torus geometries.
The dynamics is initiated in a state with a consecutive block of filled orbitals.

Quench dynamics of interacting fermions in the LLL has been recently studied
\cite{GromovPapic_geometricquenchFQH_arxiv1803}, with the quench changing the torus aspect ratio.
This type of quench is sensitive to low-energy physics (such as magnetorotons), in contrast to our
setup where the part of the many-body spectrum that is relevant is closer to the top of the spectrum
than the bottom.

Our setup is specific for the finite geometries (sphere and torus) that are common in numerical
diagonalization, so some care is required in interpreting this situation in the thermodynamic limit,
and even more care would be needed to relate to experiments.  Our initial state roughly mimics a
‘puddle’ or a ‘stripe’ of fermions in a magnetic field (a constant density of fermions in the
interior of the puddle is achieved for large enough system sizes). The state is designed to be such
that the only dynamics is due to the interactions (\ie the initial state is an eigenstate of the
magnetic field and kinetic energy parts of the Hamiltonian). While this last feature of the initial
state may seem unnatural in a continuum context, electrons confined to a stripe-shaped region will
approximately fill the orbitals in that region, and hence our initial state may be a good zeroth
approximation. Since we consider the full many-body spectrum of the LLL Hamiltonian, we are focusing
on the parameter regime in which the magnetic field is so large that the Landau level splitting is
much larger than the many-body bandwidth.

Viewed as a one-dimensional (1D) lattice system, our setup bears resemblance to inhomogeneous
quenches starting from domain-wall initial states, which is a topic of extensive interest with more
conventional 1D lattice Hamiltonians.
The perspective of the FQH interaction being treated as a 1D chain Hamiltonian pair-hopping has been
adopted previously, \eg Refs.\ \cite{LeeLeinaas_PRL2004, BergholtzNakamuraSuorsa_PhysicaE2012,
  NakamuraBergholtz_PRL2012, WangNakamura_PRB2013}.  In some cases
\cite{BergholtzNakamuraSuorsa_PhysicaE2012, NakamuraBergholtz_PRL2012, WangNakamura_PRB2013}, the
proximity to the thin torus limit allows one to truncate the longer-range terms, leading to a more
manageable or even integrable Hamiltonians.  Such a truncated Hamiltonian would not give interesting
dynamics in our case, as the pair-hopping has to be long-range enough to initiate the melting.

Viewed as a 1D chain, our interactions have the peculiarity of being neither local nor non-local in
the conventional sense.  The ``interaction range'' (length scale of the gaussian fall-off) grows as
the square root of the number of orbitals.  Thus the range grows indefinitely with system size but
the fraction of the system covered by the interaction decreases with system size.  
An interesting aspect of the $\sim\sqrt{N_{\phi}}$ growth of the interaction range is that, for
large enough system sizes, the equatorial initial state would be frozen (up to exponentially long
times) if the filling fraction were kept unchanged.  If the width of the initial block grows
linearly as $\sim N_{\phi}/3$, the interactions will not be long-ranged enough to initiate melting
at larger sizes.  However, as long as the block size is kept $\lesssim\sqrt{N_{\phi}}$, spreading
dynamics at reasonable time scales is expected at larger $N_{\phi}$ as well.

We have stressed the correlated-pair-hopping nature of our dynamics-driving terms, contrasting the
dynamical effects of these quartic terms with dynamics driven by quadratic single-particle-hopping
Hamiltonians.  Correlated-pair hopping has been rarely addressed explicitly in the quantum dynamics
literature.  Interacting Sachdev-Ye-Kitaev models (SYK$_4$), for example, contain correlated pair
hopping terms of random strengths.  The relaxation dynamics and level spacing statistics of SYK$_4$
models have been studied recently \cite{GaciaGVerbaarschot_PRD2016, EberleinSachdev_PRB2017,
  SonnerVielma_JHEP2017, McClartyHaque_arxiv2017}.  However, these models have no spatial structure,
so there are no transport/spreading effects.  Also, we are not aware of any dynamical features
explicitly due to correlated pair hopping.

More generally we may remark, as we did after equation \eqref{eq:H_tightbinding}, that any model
with one or more conserved momentum variables and two-particle interactions can be described in
momentum space and then the dynamics necessarily shows symmetric two particle hopping. Since
the momentum states usually do not have spatially varying density, this hopping is difficult to interpret in real space. Also, hopping may now occur for pairs of widely separated momenta.
Many of the phenomena we describe here are therefore specific to FQH and other chiral systems.
For example we would not expect "freezing" of wide bands of momentum states.
Nevertheless some phenomena observed here may carry over.
For example the asymptotic distribution in momentum space should depend on the conserved total initial momentum (or angular momentum).
\emph{E.g.}~for systems on the sphere it would be interesting to see under what conditions a linear
profile in angular momentum space emerges, like that observed in section \ref{sec:Lopsided}.

We have reported a (to our knowledge, first) study of the level statistics of the LLL-projected
Hamiltonians.  Level statistics is considered important for dynamics, primarily because it is
sensitive to integrability, and some classes of non-equilibrium phenomena are affected by proximity
to integrability.  For many-body systems, there is a stark difference between integrable and
non-integrable (chaotic or generic) Hamiltonians with regards to the long-time dynamics ---
integrable systems do not obey the Eigenstate Thermalization Hypothesis, and hence do not thermalize
in the same way as chaotic many-body systems \cite{RigolOlshanii_Nature2008,
  PolkovnikovSilva_RevModPhys2011, Eisert_NatPhys2015, PolkovnikovRigol_AdvPhys2016}.  In addition,
integrability affects quantum transport, leading in some cases to ballistic behavior even at finite
temperatures \cite{ZotosPrelovsek_PRB1997, HeidrichMHoneckerBrenig_EPJST2007, Prosen_PRL2011,
  SirkerPereiraAffleck_PRB2011, VasseurMoore_JSM2016, VaseurKarraschMoore_PRB2018}.  For
inhomogeneous quenches, integrability allows one to formulate a generalized hydrodynamics
\cite{BertiniColluraFagotti_PRL2016, CastroADoyon_PRX2016, ColluraDeLucaViti_PRB2018,
  VaseurKarraschMoore_PRB2018}.  The type of expansion dynamics we are considering is in spirit
reminiscent of transport; hence the presence or absence of integrability is a relevant question in
the context of the present work.  Through our analysis of level statistics, we have demonstrated
that the LLL-projected Hamiltonians are not integrable in either of the two geometries.  However,
considering the types of Wigner-Dyson statistics in various symmetry sectors, we have found that
many sectors show GOE statistics despite the broken time reversal symmetry.  These are many-body
manifestations of the mechanism proposed in Ref.\ \cite{RobnikBerry_JPA1986} --- the relevant
anti-unitary symmetry operation is not pure time reversal but a combination of time reversal and
spatial reflection.

While this has been a detailed exploration, we believe we have only scratched the surface of
fermionic LLL dynamics, and many questions remain open.  The interplay of the interaction range and
system size might give rise to nontrivial size dependencies of dynamical behaviors, which we have not
pursued.  We expect the dynamical features to depend only loosely on the filling fraction, but this
remains to be explicitly explored.  Our block initial states have been chosen in analogy to the
domain-wall literature --- a range of other initial states could be of interest and would lead to
different classes of physics.  While studying the level statistics, we have found in some cases GOE
statistics in the absence of time reversal invariance.  This calls for a more detailed analysis of
symmetries in condensed-matter models violating time reversal symmetry.

\section*{Acknowledgments}

We thank B.~Dolan, J.~Dubail, F.~Heidrich-Meisner, M.~Hermanns and  G.~Kells  for discussions.
This work was supported through SFI Principal Investigator Award 12/IA/1697.
We also wish to acknowledge the SFI/HEA Irish Centre for High-End Computing (ICHEC) for the provision of computational facilities and support through project nmphy011b.
We are grateful for the use of the code packages Hammer\cite{Hammer} and DiagHam\cite{DiagHam}, which were used for the numerical simulations.

\appendix

\section{The non-monotonicity at short times --- $V_1$ versus Coulomb
  interactions}\label{app:non-monotonic}

\begin{figure*}[tbp]
  \begin{centering}
    \includegraphics[width=.90\linewidth]{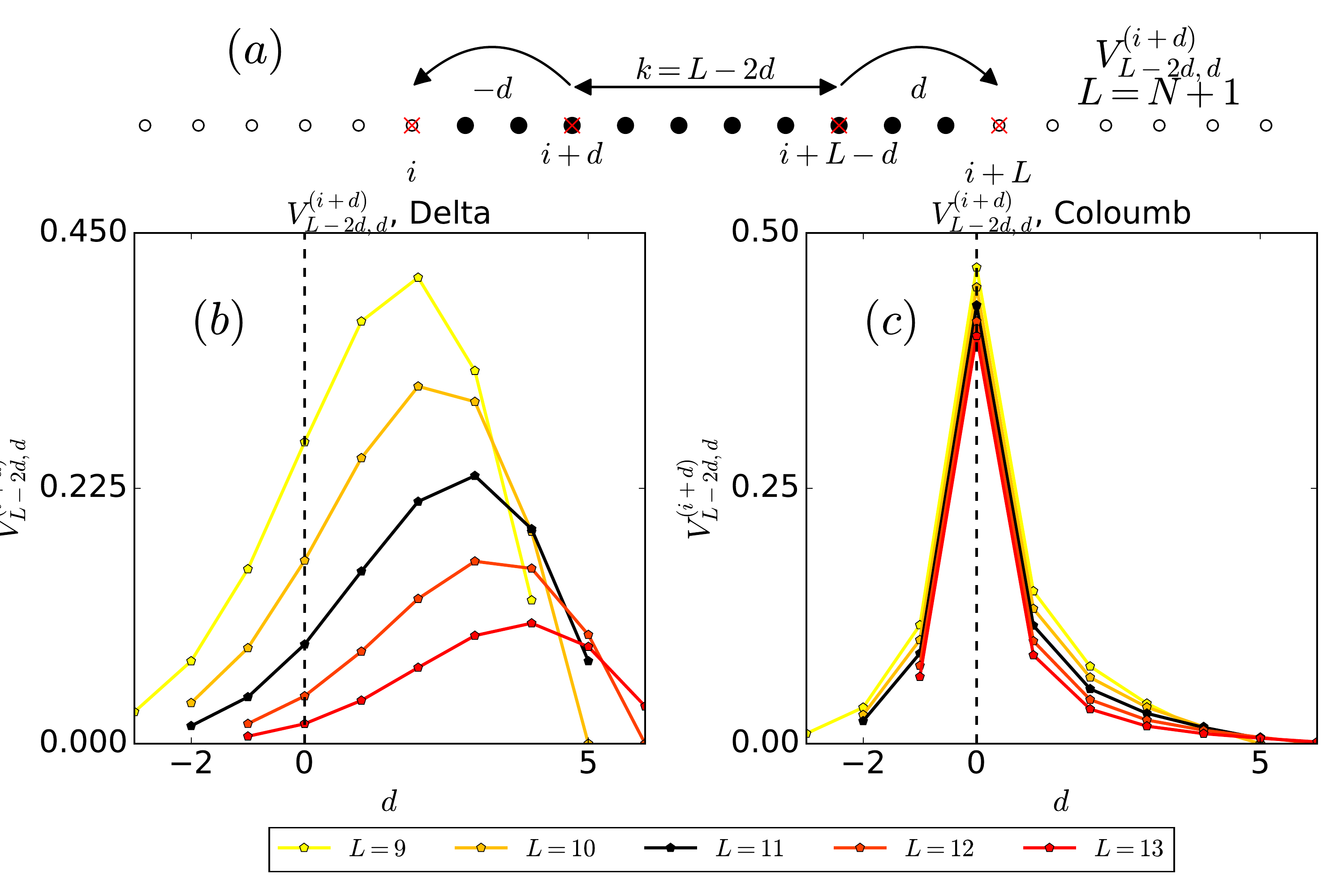}
  \end{centering}
  \caption{ Hopping elements for the $V_1$ potential and the Coulomb potential along the cut
    $V_{L-2d,d}$.  }
  \label{fig:hopping_fix_N}
\end{figure*}

In section \ref{sec:OrbEvol} it was noted that at very short times, the orbital occupancy profile is
non-monotonic.  This is a specialty of the $V_1$ potential, and is not present in the case of the
Coulomb interaction.  The non-monotonicity may be explained by noting that for a block of length
$N=L-1$ the initial de-occupation (at lowest order in time) of the orbital $d$ sites away from the
initial block boundary, is going to be driven by the hopping term $V^{i+d}_{L-2d,d}$, as shown
pictorially in \figref{fig:hopping_fix_N}(a).  The assumption here is that hopping predominantly
occurs to the first orbital outside of the block, at very short times, which is consistent with the
observations of \figref{fig:orb_dens_evol}(a).

Inspecting the values of $V^{i+d}_{L-2d,d}$ for a $N=10$ particle system (black line) in
\figref{fig:hopping_fix_N}(b), we see that the maximum is not at $d=1$, the most short ranged
allowed hopping, but rather at $d=3$.  This is consistent with the position of the non-monotonic dip
in \figref{fig:orb_dens_evol}(a). For large $N$, the position of the maximum of $V^{i+d}_{L-2d,d}$
is found to scale as $d\sim \sqrt{N}$.  This means that the distance of the dips from the edges of
the initial block should scale as $\sim \sqrt{N}$ at large $N$.

Considering instead the Coulomb interaction, \figref{fig:hopping_fix_N}(c), we see the maximum of
$V^{i+d}_{L-2d,d}$ occurs at $d=1$ ($d=0$ means no hopping) showing that this potential would not
have the same non-monotonic behaviour at early times.

\end{document}